\documentclass[11pt]{article}

\usepackage{amsmath, amssymb, latexsym, mathrsfs, pifont, float}

\newif\ifpdf
\ifx\pdfoutput\undefined
      \pdffalse
\else
      \pdftrue
\fi

\ifpdf
   \pdfcatalog { /PageMode (/UseNone)
                  /OpenAction (fitbh)
    }

  \usepackage[pdftex]{graphicx}
  \usepackage{thumbpdf}
  \usepackage[pdftex,
        colorlinks=false,
        urlcolor=rltblue,       
        filecolor=rltgreen,     
        linkcolor=rltred,       
        pdftitle={On the writhe of non-closed curves},
        pdfauthor={Eugene Starostin
                          [Eugene.Starostin@epfl.ch]},
        pdfsubject={writhe},
        pdfkeywords={writhe,
                     DNA},
        pdfpagemode=UseOutlines,
        bookmarksopen=true]{hyperref}
  \usepackage{color}
  \definecolor{rltred}{rgb}{0.75,0,0}
  \definecolor{rltgreen}{rgb}{0,0.5,0}
  \definecolor{rltblue}{rgb}{0,0,0.75}
\else
  \usepackage{graphicx}
\fi

\def\sb#1{\text{\sffamily\bfseries #1}}
\def\R{\mathbb{R}}

\newcommand{\round}{\mathop{\rm round}\nolimits}
\newcommand{\sign}{\mathop{\rm sign}\nolimits}

\newcommand{\Lk}{{\cal L}{\it k}} 
\newcommand{\Tw}{{\cal T}\hspace{-1mm}{\it w}} 
\newcommand{\Wr}{{\cal W}\hspace{-0.3mm}{\it r}} 
\newcommand{\Wy}{{\cal W}\hspace{-0.3mm}{\it y}} 
\newcommand{\Sw}{{\cal S}\hspace{-0.5mm}{\it w}} 
\newcommand{\Sq}{{\cal S}\hspace{-0.5mm}{\it q}} 

\newcommand{\Tn}{{\cal T}\hspace{-1mm}{\it n}} 

\begin{document}

\ifpdf
  \DeclareGraphicsExtensions{.pdf, .png, .jpg}
\else
  \DeclareGraphicsExtensions{.ps, .eps}
\fi



\pagestyle{myheadings}
\markright{{\rm E.L.Starostin.} On the writhe of non-closed curves}

\title{On the writhe of non-closed curves\\
{\it Preprint}}

\author{E.L. STAROSTIN  
\thanks{Permanent address: M.~V.~Keldysh Institute of Applied Mathematics,  
Russian Academy of Sciences, Miusskaya pl. 4, 125047 Moscow, Russia}
\\
{\it Institute Bernoulli, Swiss Federal Institute of Technology,}
\\
{\it CH-1015 Lausanne, Switzerland} \\
{\it Email: eugene.starostin@epfl.ch}
}

\date{}
\maketitle

\thispagestyle{empty}

\begin{abstract}
The writhe of a space curve fragment is considered for various boundary conditions.
An expression for the writhe as a function of arclength for an arbitrary space curve is obtained. 
The formula is built on the base of closing the tangent indicatrix with a geodesic. 
The corresponding closure of a curve in 3-space is explicitly constructed.
The addition rule for writhe is formulated. A relationship connecting the writhe
with the Gau\ss~integral over the open curve is presented.
The single and double regular helical shapes are examined as examples.
\vspace{4mm}

\end{abstract}

\tableofcontents


\pagebreak

\section{Introduction}

The term ``writhe'' (or ``writhing number'') was first proposed by Fuller \cite{Fuller71} for a quantity $\Wr$ that arises as a difference between the linking number $\Lk$
and the twist (or ``twisting number'') $\Tw$ of a closed ribbon in the
C\u{a}lug\u{a}reanu-White-Fuller formula
\cite{Calugareanu, White, Fuller71}
\begin{equation}
\Wr=\Lk-\Tw
\end{equation}
The sense of this very simple-looking and very famous (isn't the second a consequence of the first?) relationship is that the right-hand side, though defined for a ribbon,
depends only on its central curve.

Since $\Wr$ is a characteristic of spatial complexity of a curve, it makes this quantity worth to consider while examining lengthy physical objects.
In particular, the values of $\Wr$ have been computed in a number of works
for various models of large-scale structure of DNAs (e.g., \cite{Starostin, WB, BeardSchlick})
as well as for experimental data on these molecules \cite{FainRud}.
RNA tertiary structures and the protein folding are other neighbour areas for an application of the geometrical and topological tools developed in the DNA studies
\cite{Chechetkin}.
However, the application of the writhe is confined formally by smoothly closed shapes though quite a lot of interesting objects have their ends not joined (or joined non-smoothly). 
One should mention here the studies based on the recently developed experimental techniques of manipulation with single DNA molecules \cite{Bustamante}.
The modelling curves require an appropriately defined measure 
to characterize their arrangement in space. Therefore, it seems to be helpful to generalize the notion of the writhe onto non-closed curves and their fragments. Indeed, such trials (explicit and implicit) have been made \cite{Paraped, Marko, Vologodskii, FainRud, BouMe, BouMe00, MaggsPRL, Maggs, KutTer}.
In some cases the definitions of the writhe suggested in different works are not consistent to each other. 

The aim of this paper is to give a consistent and natural generalization of the notion of writhe to an arbitrary fragment of a curve and
to derive explicit formulas for its computation as a function of the arclength. 
The basic idea consists in a construction of the closure of the fragment under consideration in such a way that it would correspond to the closure of the tangent indicatrix by an arc of a great circle, as it has been proposed by Maggs \cite{MaggsPRL, Maggs}. In some sense, we can think then about the writhe as being locally defined (cf. \cite{Kamien}). 


\pagebreak
\section{Notation and preliminaries}

We start our consideration with a smooth non-self-intersecting curve
$A = {\bf r}(s): [0, L] \rightarrow \R^3$ of class $C^2$, $s$ being the arclength.
We assume that the segment has a natural orientation in the direction of the arc
coordinate increase. In what follows, we shall denote a reverse-oriented segment by the minus sign: $-A$. The concatenation of two segments $A$ and $B$ having, respectively, the common ending and starting point will be written as $A+B$.

A vector function ${\bf u}(s): [0, L] \rightarrow {\bf S}^2\equiv\{z\in\R^3; |z|=1\}$, ${\bf u} \in C$, may be chosen such that $s$, ${\bf u}(s) \cdot {\bf r}'(s) = 0, \forall s$; $'$~denotes the derivative with respect to $s$. Let $\epsilon > 0$ be small enough so that the ribbon $R_\epsilon = \{{\bf r}+\mu{\bf u}, -\epsilon \le \mu \le \epsilon\}$ does not cross itself.

\subsection{The twist, the writhe and the linking number}

The {\bf twisting number} (or {\bf the twist}) of the ribbon (i.e., of the pair 
$({\bf r}, {\bf u}))$ is defined by
$$
\Tw({\bf r}, {\bf u})= \frac{1}{2\pi}\int\limits_{0}^L {\bf r}' \times {\bf u} \, d{\bf u} .
$$
If ${\bf r}\in C^3$ and ${\bf r''}\neq 0$, then
the twist of a ribbon may be decomposed into the twist of the Frenet frame plus the twist of the ribbon relative to the Frenet frame \cite{Tyson}:
\begin{equation}
\Tw({\bf r}, {\bf u})= \Tw_F({\bf r}) + \frac{1}{2\pi}\int\limits_{0}^L d\phi . \label{eq:TwF}
\end{equation}
The angle $\phi = \phi(s)$ is an angle between ${\bf u}$ and the principal normal.
By the Frenet ribbon is meant a special one defined by the principal normal vector to the curve. The Frenet ribbon is defined uniquely by the space curve ${\bf r}(s)$, if ${\bf r''}\neq 0$.  The twist of the Frenet ribbon is
\begin{equation}
\Tw_F({\bf r}) = \frac{1}{2\pi}\int\limits_{0}^L \tau(s) ds , \label{eq:tau}
\end{equation}
where $\tau(s)$ is the torsion of the curve ${\bf r}(s)$. Clearly, for planar curves, $\Tw_F~=~0$. 

Now consider the smoothly closed curve: ${\bf r}(0)={\bf r}(L), {\bf r}'(0)={\bf r}'(L)$.
The quantity called {\bf the writhe} 
may be expressed as the double integral
\begin{equation}
\Wr_A=\frac{1}{4\pi}\int\limits_{0}^L\int\limits_{0}^L\frac{({\bf r}(s_1)-{\bf r}(s_2)) \cdot ({\bf t}(s_1)\times{\bf t}(s_2))}{|{\bf r}(s_1)-{\bf r}(s_2)|^3} ds_1 ds_2, \label{eq:defWr}
\end{equation}
where ${\bf t}={\bf r}'(s)$ is the tangent vector and $s_1, s_2$ are the arclengths.
The right-hand side of Eq.~(\ref{eq:defWr}) is the Gau\ss~linking integral in the singular case as being over all distinct pairs of points on one curve.
The writhe depends exclusively on the shape of the curve.

For two closed curves $A$ and $B$, $A \cap B=\emptyset$, the Gau\ss~linking integral
gives an integer-valued topological invariant
\begin{equation}
\Lk(A,B)=\frac{1}{4\pi}\int\limits_{B}\int\limits_{A}\frac{({\bf r}_A(s_1)-{\bf r}_B(s_2)) \cdot 
({\bf t}_A(s_1)\times{\bf t}_B(s_2))}{|{\bf r}_A(s_1)-{\bf r}_B(s_2)|^3} ds_1 ds_2,
\label{eq:defLk}
\end
{equation}
called {\bf the linking number}.

\subsection{The basic relations}

Two theorems by Fuller \cite{Fuller78, Aldinger} are valid. They provide the means to compute $\Wr$ efficiently.

1. Let $B={\bf r}(s)\in C^3$ be a closed oriented space curve with its tangent ${\bf r}'_B(s)$, $s$ the arclength. The tangent traces out a \emph{closed} curve $\tilde{B}(s)$ on the unit sphere which is piecewise of class $C^2$. The curve $\tilde{B}(s)$ is divided into a finite family of non-self-intersecting closed piecewise $C^2$ space curves. Each curve of this family then encloses a domain $\Omega_i$ defined so that the geodesic normal points into its interior. Let $S_B$ be the sum of the areas of these domains (the components are counted with multiplicity determined by how many times the corresponding domains are encircled by the curve). Then
\begin{equation}
\Wr({\bf r})=\frac{S_B}{2\pi} - 1 \quad \mod \ 2 . \label{eq:Fuller1}
\end{equation}

2. Let ${\bf r}_0(\theta)$ and ${\bf r}_1(\theta)$ be two closed  non-self-intersecting curves in $\R^3$ of class $C^2$, $\theta \in [0,\Theta]$ being the common regular parameter. 
Let there exist a continuous deformation 
${\bf F}(\lambda) : (\theta,\lambda)\rightarrow{\bf r}_\lambda(\theta)$,  $\lambda \in [0,1]$,  such that

\begin{enumerate}
\item[1)]
${\bf F}(0)={\bf r}_0(\theta)$ and ${\bf F}(1)={\bf r}_1(\theta)$,

\item[2)]
${\bf r}_\lambda(\theta)\in C^1$ is a non-self-intersecting curve for all $\lambda \in [0,1]$,

\item[3)]
the tangent ${\bf t}_\lambda(\theta)$ is continuous in $\lambda$,

\item[4)]
$|\angle({\bf t}_0(\theta),{\bf t}_\lambda(\theta))|<\pi,~ \forall (\theta,\lambda) \in [0, \Theta]\times[0,1]$.

\end{enumerate}

Then the difference of the writhes of the two curves is
\begin{equation}
\Wr({\bf r}_1)-\Wr({\bf r}_0)=\frac{1}{2\pi}\int\limits_0^\Theta
\frac{{\bf t}_0(\theta) \times {\bf t}_1(\theta)}
{1+{\bf t}_0(\theta) \cdot {\bf t}_1(\theta)}
\cdot \frac{d}{d\theta}\left({\bf t}_0(\theta)+{\bf t}_1(\theta)\right) d\theta .
\label{eq:Fuller2}
\end{equation}

We should stress that both theorems are applicable to \emph{closed} curves.
Nevertheless, 
the works occur (e.g., \cite{Zhou}) in which the writhe of a linear (non-closed) curve is computed by means of Eq.~(\ref{eq:Fuller2}) that leads generally to improper results. 

Let the ribbon $({\bf r}, {\bf u})$ be also closed: ${\bf u}(0)={\bf u}(L)$.
Denote by $\Lk({\bf r}-\mu{\bf u}, {\bf r}+\mu{\bf u}) \equiv \Lk({\bf r},{\bf u})$ the linking number of the two boundary curves ${\bf r}-\mu{\bf u}$ and ${\bf r}+\mu{\bf u}$. For $\epsilon$ small enough, $\Lk({\bf r},{\bf u})$ does not depend on $\epsilon$. This justifies omitting $\epsilon$ in the following. In other words, we shall be dealing with arbitrarily narrow ribbons. 

The famous C\u{a}lug\u{a}reanu-White-Fuller theorem \cite{Calugareanu, White, Fuller71} claims that
the difference of the linking and twisting numbers is the writhe:
$$
\Lk({\bf r}, {\bf u})- \Tw({\bf r}, {\bf u})= \Wr({\bf r}) .
$$ 

\subsection{The Gau\ss~integral over self-intersecting and non-smooth curves}
\label{existGauss}

We are going to show that the integral in Eq.~(\ref{eq:defWr}
) is well-defined also for curves that are 
smooth only piecewise. To demonstrate this, it is sufficient to consider a fragment of a
curve in the vicinity of the point where the tangent vector is not continuous. If the double integral
does not diverge in this point then, as a consequence, it also exists for curves crossing themselves
in a finite number of points. 

As a general case, we take the model of two curves (which may be thought of as 
two smooth fragments of the same curve) ${\bf r}_1(s_1)$ and ${\bf r}_2(s_2)$ that have one common point
$O: {\bf r}_1(0)={\bf r}_2(0)$; $s_1$ and $s_2$ are the arclengths, $s_i \in [0, l_i], i=1,2$.
Generally, ${\bf t}_1(0) \neq {\bf t}_2(0)$.
Each of the two curves is of class $C^3$ and, near $O$, it may be locally described in {\it its own} Frenet frame 
$({\bf t}_{i0}, {\bf n}_{i0}, {\bf b}_{i0}), i=1,2$, as
\begin{equation}
{\sb r}_i(s_i)=\left(s_i-\frac{\kappa_i^2}{6} s_i^3, \ 
\frac{\kappa_i}{2} s_i^2+\frac{\kappa_i'}{6} s_i^3, \ 
\frac{\kappa_i\tau_i}{6}s_i^3\right)+{\mathcal O}(s_i^4), \quad i=1,2,
\label{eq:loc1}
\end{equation}
where $\kappa_i$ and $\tau_i$ are the curvature and torsion of the $i$-th curve in the point $O$. 
(If the point $O$ is an (isolated) inflection point, it is still possible to define the local frame as the limiting Frenet
for $s_i \to 0+$.)
Differentiation of Eq.~(\ref{eq:loc1}) yields

\begin{equation}
{\sb t}_i(s_i)=\left(1-\frac{\kappa_i^2}{2} s_i^2, \ \kappa_i s_i+\frac{\kappa_i'}{2} s_i^2, \ 
\frac{\kappa_i\tau_i}{2}s_i^2\right)
+{\mathcal O}(s_i^3), \quad i=1,2.
\label{eq:loc2}
\end{equation}

Let $A=\{a_{jk}\}$ be an orthogonal matrix of orientation of the second Frenet frame with respect to the first one:
\begin{center}
\begin{tabular}{l|lll}
\noalign{\smallskip}
             & ${\bf t}_{20}$ & ${\bf n}_{20}$ & ${\bf b}_{20}$ \\
\noalign{\smallskip}\hline\noalign{\smallskip}
${\bf t}_{10}$ &              &              & \\
${\bf n}_{10}$ &              &  $a_{jk}$    & \\
${\bf b}_{10}$ &              &              & \\
\noalign{\smallskip}
\end{tabular}
\end{center}
Then the integrand of the Gau\ss~integral (Eq.~(\ref{eq:defLk}))
$$
I_{\Wr}(s_1,s_2)=\frac{({\bf r}_1(s_1)-{\bf r}_2(s_2)) \cdot ({\bf t}_1(s_1)\times{\bf t}_2(s_2))}{|{\bf r}_1(s_1)-{\bf r}_2(s_2)|^3}
$$
may be represented in the form
\begin{eqnarray}
I_{\Wr}(s_1,s_2)=\frac{({\sb r}_1(s_1)- A{\sb r}_2(s_2)) \cdot ({\sb t}_1(s_1)\times A {\sb t}_2(s_2))}{|{\sb r}_1(s_1)-A{\sb r}_2(s_2)|^3} = \nonumber \\
=\frac{A{\sb t}_2 \cdot ({\sb r}_1 \times {\sb t}_1) + {\sb t}_1 \cdot A ({\sb r}_2 \times {\sb t}_2)}{|{\sb r}_1-A{\sb r}_2|^3} . \nonumber
\end{eqnarray}
With the help of Eqs.~(\ref{eq:loc1}), (\ref{eq:loc2}), it is possible to obtain an approximation for small $s_1, s_2$
\begin{eqnarray}
I_{\Wr}(s_1,s_2)= 
\left\{ \frac12(a_{31}\kappa_1 s_1^2 + a_{13}\kappa_2 s_2^2)\right. + \nonumber \\
+\frac13 [(a_{31}\kappa_1'-a_{21}\kappa_1\tau_1)s_1^3+(a_{13}\kappa_2'-a_{12}\kappa_2\tau_2)s_2^3]+\nonumber \\
+\left. \frac12\kappa_1\kappa_2(a_{32}s_1+a_{23}s_2)s_1 s_2 + {\mathcal O}_4\right\} \times \nonumber \\
\times [s_1^2-2a_{11}s_1 s_2 +s_2^2- (a_{21}\kappa_1 s_1 + a_{12}\kappa_2 s_2)s_1 s_2 + {\mathcal O}_4]^{-\frac32},
\label{eq:loc3}
\end{eqnarray}
where the terms of the 4-th order and higher are denoted by $${\mathcal O}_4 = \sum\limits_{l=0}^{4}{\mathcal O}(s_1^l s_2^{4-l})$$.

Consider first the case when $a_{11}^2 \neq 1$. 
It is convenient to introduce the new variables $\rho, \phi$ ($\rho \ge 0$, $\phi \in [0, \frac{\pi}{2}]$) such that
\begin{eqnarray}
s_1 = \frac{\sqrt{2}}{2}\rho(\sqrt{1-a_{11}}\cos\phi+\sqrt{1+a_{11}}\sin\phi), \nonumber \\
s_2 = \frac{\sqrt{2}}{2}\rho(\sqrt{1+a_{11}}\sin\phi-\sqrt{1-a_{11}}\cos\phi). \nonumber
\end{eqnarray}
Then, with the use of Eq.~(\ref{eq:loc3}), we have
\begin{eqnarray}
\int\int I_{\Wr}(s_1,s_2) ds_1 ds_2 = 
\int\int \hat{I}_{\Wr}(\rho,\phi) \sqrt{1-a_{11}^2} \ \rho \ d \rho \ d \phi = \nonumber \\
=\frac1{1-a_{11}^2}\int\int\frac{\Phi_0(\phi)\rho^3+\Phi_1(\phi)\rho^4 + {\mathcal O}(\rho^5)}{\rho^3[1+\Psi_1(\phi)\rho+{\mathcal O}(\rho^2)]^{\frac32}}\ d\rho \ d\phi . \nonumber
\end{eqnarray}

As $\rho \to 0$, the integrand of the last integral approaches
$$
\Phi_0(\phi)=\frac14[(a_{31}\kappa_1+a_{13}\kappa_2)(1-a_{11}\cos{2\phi})+\sqrt{1-a_{11}^2}(a_{31}\kappa_1-a_{13}\kappa_2)\sin{2\phi}].
$$
It vanishes only if $a_{31}\kappa_1=0$ and $a_{13}\kappa_2=0$. This happens when either 1) the both curves have $O$ as an inflection point or 2) one curve has an inflection point and its tangent lies in the tangent plane of the other one or 3) the tangent planes of both curves coincide.

Now let $a_{11} = -1$. The meaning of this is that the curves ${\bf r}_1$ and ${\bf r}_2$ form a curve having a continuous tangent.
We introduce the angle $\chi$ which parametrizes the relative orientation of the Frenet frames at $O$ (or one can interpret $\chi$
as a discontinuity angle between principal normals). Then we have for the entries of the orientation matrix:
$a_{12} = a_{21} = a_{13} = a_{31} = 0$, $a_{22} = - a_{33} = \cos\chi$, $a_{23} = a_{32} = \sin\chi$. 
Eq.~(\ref{eq:loc3}) takes the form
$$
I_{\Wr}(s_1,s_2)= 
\frac{\frac12\kappa_1 \kappa_2 \sin\chi (s_1 + s_2)s_1s_2 + {\mathcal O}_4}
{[(s_1 + s_2)^2 + {\mathcal O}_4]^{-\frac32}} .
$$

Let us redefine the variables $\rho$ and $\phi$: $s_1 = \rho\sin\phi$, $s_2 = \rho\cos\phi$ ($\rho \ge 0$, $\phi \in [0, \frac{\pi}{2}]$), then
\begin{eqnarray}
\int\int I_{\Wr}(s_1,s_2) ds_1 ds_2 = \int\int \hat{I}_{\Wr}(\rho,\phi) \rho \ d \rho \ d \phi = \nonumber \\
=\int\int\frac{\frac12\kappa_1 \kappa_2(\sin\phi + \cos\phi)\sin\phi\cos\phi\rho^4 + {\mathcal O}(\rho^5)}{\rho^3[(\sin\phi + \cos\phi)^2 +{\mathcal O}(\rho^2)]^{\frac32}}\ d\rho \ d\phi , \nonumber
\end{eqnarray}
from where we can see that the integral exists.

The case $a_{11} = 1$ corresponds to the cusp points and requires a special consideration which is beyond the scope of the present paper.
Therefore we exclude cusps from the following analysis.


\section{A periodic curve}

On the one hand, the treatment of the periodic curve case is the most resembling to that of the closed configuration. On the other hand, the periodic curves are often
met in various physical and, in particular, biomechanical models. For example, the linear helices and interwound structures are common in the DNA modelling (e.g., 
see \cite{Tanaka}).

\subsection{A formula for the writhe}

Let ${\bf r}_A(s)$ be periodic in space: ${\bf r}_A(s+L_A)={\bf r}_{AL}+{\bf r}_A(s)$,
$L_A$ the period, ${\bf r}_{AL} = {\bf const}$. For the tangent ${\bf t}_A(s)={\bf r}'_A(s)$, it implies
${\bf t}_A(s+L_A)={\bf t}_A(s)$.
Denote ${\bf t}_{A0}={\bf t}_{A}(0)$ and ${\bf t}_{A1}={\bf t}_{A}(L_A)$.
Consider the segment $A: 0 \le s \le L_A$. The tangent indicatrix is a \emph{closed} curve which sweeps out an area $S_A$ on the unit sphere ${\bf S}^2$ (in the same sense as in the Fuller first theorem). Now construct a ribbon $R_A$ for the segment $A$ in the same manner as it is done in the proof of the Fuller first theorem \cite{Aldinger}. 
Generally, the ribbon $R_A$ is determined by the unit principal normal vector ${\bf n}_A(s)$. However, we should account for the possible inflection points of $A$ with the discontinuous normals. Because of this, in the vicinities of the inflection points the ribbon $R_A$ is to be arbitrary modified to make it continuous with a new modified generating vector ${\bf u}_A$.
Due to the periodicity property, 
${\bf u}_{A0} \equiv {\bf u}_{A}(0) = {\bf u}_{A}(L_A) \equiv {\bf u}_{A1}$.
For the sake of simplicity, we shall be also assuming, here and further, that
${\bf u}_{A0} = {\bf n}_{A}(0)$ and ${\bf u}_{A1} = {\bf n}_{A}(L_A)$.

The twist of $R_A$, which is well defined for non-closed ribbons as well, satisfies the equation
\begin{equation}
\Tw({\bf r}_A, {\bf u}_A)+\frac{S_A}{2\pi} = 0 \quad \mod \ 1 \label{eq:TS}
\end{equation}
(the above follows from the proof of the Fuller first theorem which, in turn, is a consequence of the Gau\ss-Bonnet theorem).

At each point of the ribbon $A$ we can define an orthonormal frame
$\{{\bf t}_A(s), {\bf u}_A(s), {\bf t}_A(s) \times {\bf u}_A(s)\}$,
which is an element of $SO(3)$. 
Thus, the ribbon $A$ corresponds to a closed loop $\hat{A}$ in the space of orientations $SO(3)$ \cite{Penrose}.
Following Hannay \cite{Hannay}, we define an integer number $N \in \mathbb{Z}_2$ which is zero if $\hat{A}$
may be continuously deformed to a point (i.e., if it is contractible) and 
$N=1$ otherwise. Then Eq.~(\ref{eq:TS}
) is a consequence of the Hannay for\-mu\-la
$$
\Tw({\bf r}_A, {\bf u}_A)+\frac{S_A}{2\pi} = N \quad \mod \ 2 . 
$$

Now build a planar non-self-intersecting segment $C: {\bf r}_{C}(s) \in C^3$, for $0 \le s \le L_C$, that closes the segment $A$. That means that 
${\bf r}_C(0) = {\bf r}_A(L_A)$, ${\bf r}_C(L_C) = {\bf r}_A(0)$, 
${\bf t}_C(0) = {\bf t}_{A1}$, ${\bf t}_C(L_C) = {\bf t}_{A0}$
(${\bf t}_C\equiv{\bf r}'_C(s)$) and
$C$ lies in the plane determined by ${\bf t}_{A0}$ and
${\bf r}_{AL}$ (if ${\bf t}_{A0} \parallel {\bf r}_{AL}$, then we choose the plane spanned by ${\bf t}_{A0}$ and ${\bf t}'_{A0}$).
The closing segment $C$ may be chosen such that its $\epsilon$-vicinity does not cross the ribbon $R_A$.
In addition, we require that the principal normal at the beginning and at the end points be well defined
and have the same direction:
${\bf n}_C(0)={\bf n}_C(L_C)$.
Then the number of inflection points of $C$ is even.

There exists a ribbon $({\bf r}_C(s), {\bf v}_C(s))$ lying entirely in the plane of $C$
with its ends defined by the normal vectors ${\bf v}_C(0)={\bf n}_C(0)$ and ${\bf v}_C(L_C)={\bf n}_C(L_C)$.
Clearly, its twist is zero.
(In the case of no inflection points, this ribbon coincides with the Frenet ribbon.)
Construct another ribbon on the curve $C$ with a vector ${\bf u}_C(s)$,
${\bf u}_C(s), \ {\bf u}_C(s) \cdot {\bf t}_C(s) \equiv 0$,
which differs from the vector ${\bf v}_C(s)$ by a constant angle:
${\bf u}_C(s) \cdot {\bf v}_C(s) = \cos\phi_0 = const$, and ${\bf u}_C(0)={\bf u}_{A1}$.
It is evident that  ${\bf u}_C(L_C)={\bf u}_C(0)$,
${\bf u}_C(L_C)={\bf u}_{A0}$ and $\Tw({\bf r}_C, {\bf u}_C)=0$.

Thus, joining the (oriented) ribbons $A: ({\bf r}_A, {\bf u}_A)$ and
$C: ({\bf r}_C, {\bf u}_C)$ gives a continuous closed ribbon which we shall denote
by $A+C$.
The linking number of a closed ribbon is an integer and, by applying the C\u{a}lug\u{a}reanu-White-Fuller relation to $A+C$, we have
\begin{equation}
\Lk_{A+C}=\Wr_{A+C}+\Tw({\bf r}_A, {\bf u}_A)+\Tw({\bf r}_C, {\bf u}_C) . \label{eq:LWTT}
\end{equation}

From Eq. (\ref{eq:LWTT}) we obtain $\Wr_{A+C}=\Lk_{A+C}-\Tw({\bf r}_A, {\bf u}_A)$
and, taking into account Eq.~(\ref{eq:TS}),
\begin{equation}
\Wr_{A+C}=\frac{S_A}{2\pi} \quad \mod \ 1 . \label{eq:WS}
\end{equation}

In \cite{comment} it was shown that the last relation may be refined
as follows
\begin{equation}
\Wr_{A+C}=\frac{S_A}{2\pi} + \Tn_C + 1 \quad \mod \ 2 , \label{eq:Wr2}
\end{equation}
where $\Tn_C$ is the turning number of the closing curve.

Since $\Tn_C$ is integer,
the fractional part of the writhe does not depend on the closing curve $C$ 
and we may write
$$
\Wr_{A}=\frac{S_A}{2\pi} \quad \mod \ 1 . 
$$

We see that in the case of a periodic curve the same formula is valid for the writhe of a period as in the case of a smoothly closed loop.

Note that the tangent indicatrix of the joined ribbon differs from that of $A$
by an appendage corresponding to the planar segment $C$. 
This appendage has an area of $2 \pi \Tn_C$, hence, 
it does not contribute to the total spherical area of the domain enclosed by $A+C$,
counted modulo $2 \pi$. In other words, $S_{A+C}=S_A \ \mod 2\pi$. 

If the writhe of a period is to be calculated by using the double integral
formula of Eq. (\ref{eq:defWr}), then
at first the segment of the period should be closed by an additional planar segment
as described above and after that the formula Eq. (\ref{eq:defWr}) should be applied to the entire closed loop, though the double integral in Eq. (\ref{eq:defWr}) is also well defined for open segments.

Another way to compute the value of the writhe $\Wr_A$ is 
the application of Eq.~(\ref{eq:LWTT}):
\begin{equation}
\Wr_A=\Lk_{A+C}-\Tw({\bf r}_A, {\bf u}_A) , \label{eq:WLT}
\end{equation}
which is valid
for any appropriate closing segment built as described above. Note also that Eq. (\ref{eq:WLT}) is an \emph{exact} formula.
With the help of the relation between the linking number and the contractibility number \cite{comment}, 
Eq. (\ref{eq:WLT}) may be written in the form
\begin{equation}
\Wr_A=N+\Tn_C+1-\frac{1}{2\pi}\int\limits_{0}^{L_A} {\bf r}'_A \times {\bf u}_A \cdot d{\bf u}_A
\quad \mod \ 2 . \label{eq:Wrn2}
\end{equation}

Eq. (\ref{eq:Wrn2}) may be useful since one often needs to know only the fractional part of the writhe. 


\subsection{Example}

\begin{figure}
[htbp]
\begin{center}
\includegraphics[height=8cm]{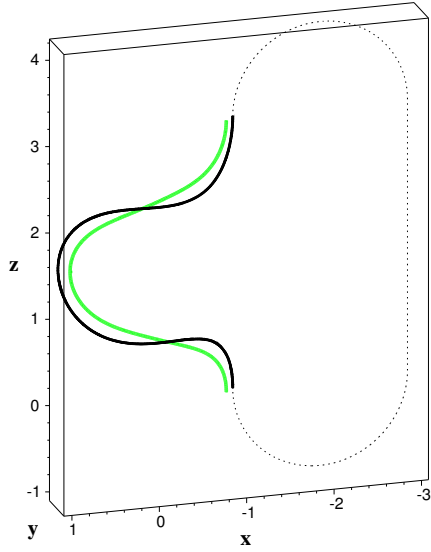}
\end{center}
\caption{One period of the curve of 
Eq.~(\ref{eq:r8}) 
(solid line), its planar closure (dashed line) and the second edge of the Frenet ribbon (thick line).}
\label{eq:example_periodic8}
\end{figure}

\begin{figure}
[htbp]   
\begin{center}
\includegraphics[height=8cm]{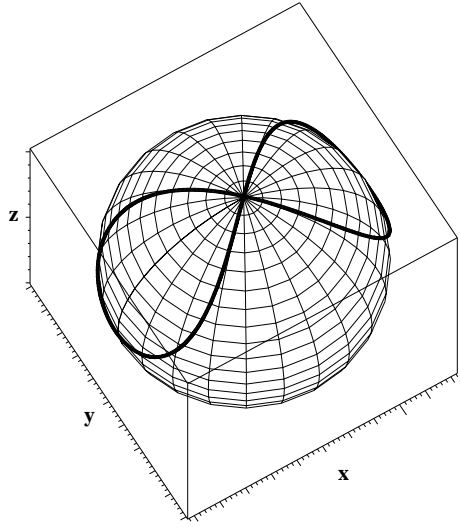}
\end{center}
\caption{The tangent indicatrix (Eq.~(\ref{eq:it8})).}
\label{eq:indicatrix8}
\end{figure}

Consider a $2\pi$-periodic spatial curve
\begin{equation}
{\bf r}=(-\cos s, -\frac{1}{4}\cos 2s, \frac{s}{2}+\frac{1}{4}\sin 2s) \label{eq:r8}
\end{equation}
(Fig.~\ref{eq:example_periodic8}). Its tangent vector is
\begin{equation}
{\bf t}=(\sin s, \frac{1}{2}\sin 2s, \cos^2 s)   \label{eq:it8}
\end{equation}
and the tangent indicatrix is a 8-shaped self-intersecting curve consisting of the two identical loops (Fig.~\ref{eq:indicatrix8}). Let $S_1$ be the spherical area inside one loop,
then the complementary area of the other loop equals $4\pi-S_1$. Thus, the sum of the areas of the two domains defined as in the Fuller first theorem, amounts to $4\pi$, and, according to Eq. (\ref{eq:Wr2}), $\Wr=0~\mod~2$.

The last conclusion is evidently true for every periodic curve, 
the tangent indicatrix of which takes a symmetric form of the figure-8 with the identical loops.
(In fact, the loops need only have equal areas.)

This result may be easily verified (and refined) by applying Eq.~(\ref{eq:WLT}). To this end, we first construct a Frenet ribbon with the principal normal vector 
$$
{\bf n}= \frac{1}{\kappa}
\frac{d {\bf t}}{d s} = \frac{1}{\kappa}(\cos s, \cos 2s, -\sin 2s) ,
$$
the curvature $\kappa=\sqrt{1+\cos^2s}$. One edge of the Frenet ribbon ${\bf r}+\epsilon{\bf n}$
is shown in Fig.~\ref{eq:example_periodic8} as a thick line. The one-period segment may be closed with a planar curve as described above and an additional segment of the ribbon is to be built to close the Frenet ribbon of the initial curve. It is easy to check that the linking of the two edges for the entire closed loop is zero. Now compute the twist of the open Frenet ribbon by 
using Eq.~(\ref{eq:tau}) into which the torsion $\tau=\frac{1}{\kappa^2} {\bf r}'
{\bf r}''{\bf r}''' = -\sin s \ \frac{\cos^2 s +2}{\cos^2 s +1}$ is substituted.
Clearly,
$$
\Tw_F({\bf r}) = \frac{1}{2\pi}\int\limits_{0}^{2\pi} \tau(s) ds =0 ,
$$
and by Eq. (\ref{eq:WLT}) we obtain finally $\Wr=0$.


\section{An open segment of a curve with the parallel initial and terminal tangents}

\begin{figure}[htbp]
\begin{center}
\includegraphics[height=8cm]{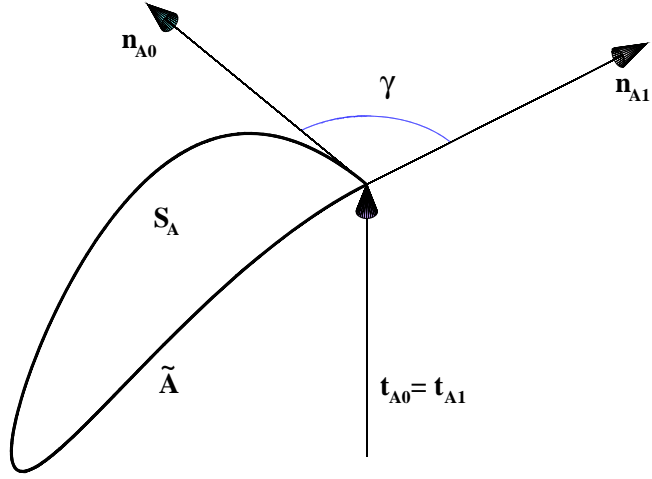}
\end{center}
\caption{The closed tangent indicatrix $\tilde{A}$ of the segment $A$.}
\label{eq:loop_gamma}
\end{figure}

We again deal with the segment $A$ of a spatial curve ${\bf r}_A(s), 0 \le s \le L_A$. 
We now make no assumptions on the behaviour of this curve outside the segment $A$. 
Only one additional condition is to be satisfied: 
the tangent vector at the beginning is the same as one at the ending point: 
${\bf t}_{A0} \equiv {\bf t}_A(0) = {\bf t}_A(L_A) \equiv {\bf t}_{A1}$.
Then the tangent indicatrix is again a \emph{closed} curve sweeping out the area
$S_A$. We construct the ribbon $R_A$ for the segment $A$ by using the unit vector ${\bf u}_A(s), \ {\bf u}_A(s) \cdot {\bf t}_A(s) \equiv 0$, in the same manner as in the case of the periodic curve. The difference now is that, generally, 
${\bf u}_{A0} \equiv {\bf u}_A(0) \neq {\bf u}_A(L_A) \equiv {\bf u}_{A1}$.
We denote by $\gamma$ the angle from ${\bf u}_{A1}$ to ${\bf u}_{A0}$ (actually,
from ${\bf n}_{A1}$ to ${\bf n}_{A0}$)
(Fig.~ \ref{eq:loop_gamma}).
Note that both vectors lie in the plane orthogonal to ${\bf t}_{A0} = {\bf t}_{A1}$.

Due to the Gau\ss-Bonnet formula \cite{Spivak},
\begin{equation}
S_A+2\pi \Tw({\bf r}_A, {\bf u}_A)+\gamma = 0 \quad \mod \ 2\pi . \label{eq:GB}
\end{equation}

The next step is to build a planar non-self-intersecting segment $C: {\bf r}_C(s) \in C^3$, for $0\le s \le L_C$, to close the segment $A$. $C$ lies in the plane spanned by ${\bf t}_{A0}$ and ${\bf r}_{AL}$ (or ${\bf t}'_A(0)$, if ${\bf t}_{A0} \parallel {\bf r}_{AL}$) and  
${\bf r}_C(0) = {\bf r}_A(L_A)$, 
${\bf r}_C(L_C) = {\bf r}_A(0)$, ${\bf t}_C(0) = {\bf t}_{A1}$, 
${\bf t}_C(L_C) = {\bf t}_{A0}$. We can also require that $C$ have
an even number of inflection points, all of them being interior.
In addition, $C$ is such that it is separated from the ribbon $R_A$ by the distance more than $\epsilon$.

A ribbon on $C$ exists that lies in the same plane as $C$ and with its end vectors
coinciding with
${\bf n}_C(0)={\bf n}_C(L_C)$. Denote its generating vector by ${\bf v}_C(s)$.
We build a ribbon on the curve $C$ by using a vector ${\bf u}_C(s)$,
${\bf u}_C(s)\cdot{\bf t}_C(s)=0$, which is the vector ${\bf v}_C(s)$ turned through a constant angle $\phi_0$:
${\bf u}_C(s) \cdot {\bf v}_C(s) = \cos\phi_0=const$. Choose $\phi_0$ such that ${\bf u}_C(0)={\bf u}_{A1}$.
By the construction of $C$, ${\bf u}_C(L_C)={\bf u}_C(0)$ and the ribbon $C$
has zero twist: $\Tw({\bf r}_C, {\bf u}_C)=0$.

We see that the joined ribbon $A+C$ may have only one point of discontinuity, namely, the starting point of $A$ (corresponding to the terminal point of $C$).
At this point the twist angle between ${\bf u}_C(L_C)$ and ${\bf u}_{A0}$ is exactly $\gamma$.

The following relation is valid, according to \cite{Maddocks}, for the discontinuous ribbon $A+C$ 
(see also Appendix~A for a different proof)
\begin{equation}
2\pi(\Tw_{A+C}+\Wr_{A+C})+\gamma=0 \quad \mod \ 2\pi . \label{eq:Mad}
\end{equation}

The sum $\Tw_{A+C}+\Wr_{A+C}$, which is not integer, may be interpreted as
the linking number  of \emph{a cord}. The latter is a generalization of a notion of the ribbon and was introduced by Fuller in \cite{Fuller78}. 

Since the twist is additive, $\Tw_{A+C}=\Tw_A+\Tw_C$, $\Tw_A\equiv\Tw({\bf r}_A,
{\bf u}_A)$, $\Tw_C\equiv\Tw({\bf r}_C, {\bf u}_C)=0$.

Combining Eq. (\ref{eq:GB}) and Eq. (\ref{eq:Mad}) leads to
\begin{equation}
\Wr_{A+C}=\frac{S_A}{2\pi} \quad \mod \ 1 .
\end{equation}

By the argument similar to that used in the case of the periodic curve it may be shown that
\begin{equation}
\Wr_{A+C}=\frac{S_A}{2\pi} + \Tn_C + 1 \quad \mod \ 2 . \label{eq:WRAS}
\end{equation}

Again, the tangent indicatrix of $C$ does not change the area of the domain 
enclosed by $A+C$, counted $\mod 2\pi$, and we conclude that the fractional part of the writhe 
does not depend on the particular shape of the closing curve constructed under the above conditions:
\begin{equation}
\Wr_{A}=\frac{S_A}{2\pi} \quad \mod \ 1 .
\end{equation}

The fractional component of $\Wr$ may be also computed by means of Eq.~(\ref{eq:Mad}):
\begin{equation}
\Wr_A=-\Tw_A-\frac{\gamma}{2\pi} \quad \mod \ 1 . \label{eq:WTgam}
\end{equation}
Note that Eq. (\ref{eq:WTgam}) does not refer to the closing segment. All what is necessary to know is the twist for the original segment $A$ and the angle $\gamma$ between the normals ${\bf n}_{A0}$ and ${\bf n}_{A1}$.
An analogue to Eq.~(\ref{eq:WTgam}) was used in the analysis of the elongation
of a supercoiled DNA molecule carried out by Bouchiat and M\'{e}zard  \cite{BouMe,
BouMe00} (though their angle $\chi$ is measured in the opposite direction to~$\gamma$).


\section{An arbitrary open segment}

\subsection{The general case}
\label{arbitrary}

The most general case takes place if we dismiss the equality condition 
of the tangent vectors at both ends of the spatial curve segment $A: {\bf r}_A(s), 0 \le s \le L_A$. 
In other words, the tangent indicatrix has not to be closed any more. Nevertheless, we can construct the ribbon $R_A: ({\bf r}_A,
{\bf u}_A)$ in exactly the same manner as we did it previously. The orientations of the generating vector ${\bf u}_A$ at the ends may be different.

In order to get a measure for the writhe of $A$ consistent with the above considered cases we choose to close the tangent indicatrix with a geodesic \cite{MaggsPRL, Maggs}. This choice is natural and it is supported by treatment of analogous problems in optics and quantum mechanics \cite{Samuel}. (In the generic case ${\bf t}_{A0} \neq \pm {\bf t}_{A1}$, there are two possible geodesics, we take one of them; the case 
${\bf t}_{A0} = - {\bf t}_{A1}$ will be discussed later.)

Let $G$ be a planar segment ${\bf r}_G(s), 0\le s \le L_G$, in the plane determined by ${\bf t}_{A0}, {\bf t}_{A1}$ such that ${\bf r}_G(0)={\bf r}_A(L_A)$ and ${\bf r}'_G(0)={\bf t}_{A1}$, ${\bf r}'_G(L_G)={\bf t}_{A0}$.
We may always require that $G$ have no inflection points.

We build a ribbon $R_G$ based on the segment $G$ and the vector ${\bf u}_G(s), 0 \le s \le L_G$, which is the principal normal ${\bf n}_G(s)$ turned around the tangent ${\bf t}_G$ through the constant angle $-\gamma_1$ such that ${\bf u}_{G0}\equiv{\bf u}_G(0)={\bf u}_{A1}$. Therefore, the two ribbons $R_A$ and $R_G$ are continuously glued to produce the joined ribbon $R_{A+G}$.

We now can see that the assumption of the previous case is satisfied for the joined curve $A+G$:
it has the same tangents at the ends.
By applying Eq.~(\ref{eq:WRAS}) to $A+G$, we obtain
\begin{equation}
\Wr_{A+G}=\frac{S_{A+G}}{2\pi} + \Tn_C +1 \quad \mod \ 2 . \label{eq:WRAS2}
\end{equation}

It should be noted that although Eq.~(\ref{eq:WRAS2}) is valid $\mod~2$, actually we may define and compute only fractional part of the writhe because of an arbitrariness of one of two closing geodesics chosen.
To put this another way, we are able to determine the area only $\mod~2\pi$. Therefore, the writhe of an open segment may be determined by the following relation:
\begin{equation}
\Wr_{A}=\frac{S_{A+G}}{2\pi} \quad \mod \ 1 . \label{eq:WRAS1}
\end{equation}

Instead of Eq.~(\ref{eq:WTgam}), we come to
\begin{equation}
\Wr_{A+G}=-\Tw_{A+G}-\frac{\gamma}{2\pi} \quad \mod \ 1 , \label{eq:WTgam2}
\end{equation}
where $\Tw_{A+G}=\Tw_A+\Tw_G$. But $\Tw_G=\Tw({\bf r}_G, {\bf u}_G)=0$ by the same argument as for the closing segment for a periodic curve.

\begin{figure}[htbp]
\begin{center}
\includegraphics[height=8cm]{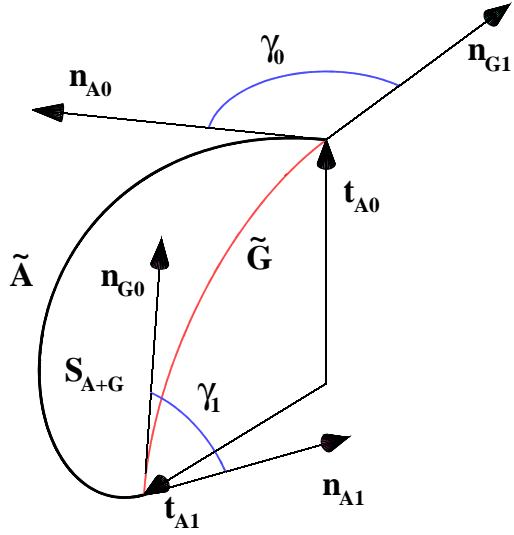}
\end{center}
\caption{The tangent indicatrix $\tilde{A}$ of the segment $A$ closed with a geodesic~$\tilde{G}$.}
\label{eq:geodesic}
\end{figure}

The angle $\gamma$ is the angle measured from ${\bf u}_{G1}\equiv{\bf u}_G(L_G)$ to ${\bf u}_{A0}$.
It may be represented as the sum $\gamma=\gamma_1+\gamma_0$, where $\gamma_0$ is an angle from the normal ${\bf n}_{G1}$ to ${\bf u}_{A0}(={\bf n}_{A0})$ 
(Fig.~\ref{eq:geodesic}).

Finally, we arrive at
\begin{equation}
\Wr_A=-\Tw_A-\frac{\gamma_1+\gamma_0}{2\pi} \quad \mod \ 1 , \label{eq:WTgams}
\end{equation}
where the angles $\gamma_1$ and $\gamma_0$ are determined by the equations
$$
\cos\gamma_1 = {\bf n}_{A1} \cdot {\bf n}_{G0}, \quad
\sin\gamma_1 = ({\bf n}_{A1} \times {\bf n}_{G0})\cdot {\bf t}_{A1},
$$
$$
\cos\gamma_0 = {\bf n}_{A0} \cdot {\bf n}_{G1}, \quad
\sin\gamma_0 = ({\bf n}_{G1} \times {\bf n}_{A0})\cdot {\bf t}_{A0}.
$$

The normal vectors ${\bf n}_{G0}$ and ${\bf n}_{G1}$ may be easily expressed as functions of the  initial and terminal tangents of $A$. Indeed, the vector ${\bf n}_{G0}$ lies in the plane spanned by the vectors ${\bf t}_{A0}$ and ${\bf t}_{A1}$. Also, ${\bf n}_{G0}\cdot{\bf t}_{A1} = 0$ and ${\bf n}_{G0} \cdot {\bf t}_{A0} > 0$. Hence,
$$
{\bf n}_{G0}={{\bf t}_{A0} - ({\bf t}_{A0} \cdot {\bf t}_{A1}) {\bf t}_{A1}
\over \sqrt{1-({\bf t}_{A0} \cdot {\bf t}_{A1})^2}} .
$$
Similarly, the vector ${\bf n}_{G1}$ lies in the same plane and ${\bf n}_{G1}\cdot{\bf t}_{A0} = 0$ and ${\bf n}_{G1} \cdot {\bf t}_{A1} < 0$. Therefore,
$$
{\bf n}_{G1}=-{{\bf t}_{A1} - ({\bf t}_{A0} \cdot {\bf t}_{A1}) {\bf t}_{A0}
\over \sqrt{1-({\bf t}_{A0} \cdot {\bf t}_{A1})^2}} .
$$

The angles $\gamma_1$ and $\gamma_0$ can be found from their trigonometrical functions
\begin{eqnarray}
\cos\gamma_1={{\bf n}_{A1} \cdot {\bf t}_{A0} \over 
\sqrt{1-({\bf t}_{A0} \cdot {\bf t}_{A1})^2}} , \quad
\sin\gamma_1={{\bf b}_{A1} \cdot {\bf t}_{A0} \over 
\sqrt{1-({\bf t}_{A0} \cdot {\bf t}_{A1})^2}} , \label{eq:CSG1} \\
\cos\gamma_0={-{\bf n}_{A0} \cdot {\bf t}_{A1} \over 
\sqrt{1-({\bf t}_{A0} \cdot {\bf t}_{A1})^2}} , \quad
\sin\gamma_0={{\bf b}_{A0} \cdot {\bf t}_{A1} \over 
\sqrt{1-({\bf t}_{A0} \cdot {\bf t}_{A1})^2}} , \label{eq:CSG2}
\end{eqnarray}
where ${\bf b}_A(s)={\bf t}_A(s) \times {\bf n}_A(s)$
is the binormal vector and ${\bf b}_{A0}\equiv{\bf b}_A(0)$,
${\bf b}_{A1}\equiv{\bf b}_A(L_A)$.

We can conclude that Eq.~(\ref{eq:WTgams}), together with Eqs.~(\ref{eq:CSG1}), (\ref{eq:CSG2}), provides
a means to compute the fractional part of the writhe for an (almost) arbitrary curve with open ends.

{\bf Remark 1.} It follows from the above that the writhe of a curve segment (closed or open) whose tangent indicatrix is geodesic and such that ${\bf t}_{A0}+{\bf t}_{A1}\neq 0$, is an integer. In other words, the writhe of any planar curve is always integer.

Consider now the case ${\bf t}_{A0} = - {\bf t}_{A1}$. The closing geodesic $G$ is not determined uniquely then. 
If we examine the behaviour of the tangent indicatrix in the vicinities $s=0$ and $s=L_A$ (also paying attention to the neighbour curves on ${\bf S}^2$ close to the critical one), then
we see that the plane of the closing geodesics may rotate through $\sim\pi$ as the length of the segment changes so that the critical point ${\bf t}_{A0} = - {\bf t}_{A1}$ is passed. In mechanics terms, this phenomenon may be called flipping of the closing segment. It is the critical point where the choice of one geodesics based on the continuity argument is no more valid.

Speaking more strictly, the value of the writhe for a segment with the oppositely directed ends is not determined. It may be ascribed a value which is average of the two limits taken as the length of the segment is pre-critical and post-critical. That means that the great circle plane for the closing
geodesics in the critical point has to be taken orthogonal to the limiting positions both of the great circle planes chosen for the growing segment 
$[\epsilon, L_A-\epsilon]$ and  the decreasing one $[-\epsilon, L_A+\epsilon]$
as $\epsilon \to 0$. (We assume that the definition segment
for the curve $A$ may be infinitesimally extended in both directions.)  
Formally, this choice is a plane spanned by 
${\bf t}_{A0}, {\bf n}_{A0}+ {\bf n}_{A1}$. For such a closure, 
$\gamma_1 + \gamma_0 = \pi$ and
Eq.~(\ref{eq:WTgams}) becomes
$$
\Wr_A=-\Tw_A -\frac{1}{2} \quad \mod \ 1 , 
$$
while Eq.~(\ref{eq:WRAS1}) does not change (as usual, by $S_{A+G}$ is meant the area swept out by the closed curve $A+G$ defined as above).
If, in addition, ${\bf n}_{A0} + {\bf n}_{A1}=0$, then a plane 
spanned by ${\bf t}_{A0}$ and ${\bf b}_{A0}$ has to be chosen.

{\bf Remark 2.} In the above consideration we have used the ribbon generally
based on the principal normal (cf. \cite{Aldinger}), though any other continuous ribbon may be taken to obtain essentially the same formula for writhe (of course, the specific expressions
for the angles $\gamma_1$ and $\gamma_0$ should be appropriately modified).

In particular, the fractional part of writhe equals the twist of a special ribbon such that
its generating normal vectors at the ends
have they property that they could be transformed to each other by a parallel transport 
along the closing geodesics (i.e. $\gamma_1 + \gamma_0 = 0 \ \mod 2\pi$).

\subsection{A broken curve}
\label{broken}

The above approach may be naturally extended to a sequence of disjoint segments.
Let $A=\sum\limits_{i=1}^{n} A^{(i)}$ be a set of $n$ continuous fragments. Each $A^{(i)}$
is oriented so that $A_0^{(i)}$ and $A_1^{(i)}$ be its initial and terminal points, respectively.
Based on $A^{(i)}$, a ribbon $R_A^{(i)}$ may be built as it was done for a single piece of curve.
We also construct $n$ additional pieces that tie the ending point of the $i$-th fragment to
the initial point of the consequent one. We identify formally the point $A_0^{(n+1)}$ with $A_0^{(1)}$
to make the entire curve closed.
The connecting parts are built in exactly the same way as the closure of a single segment was made in the previous subsection. Thus, we can repeat our arguments to obtain
\begin{equation}
\Wr_A=-\sum\limits_{i=1}^{n}\Tw_i-\frac{1}{2\pi}\sum\limits_{i=1}^{n}(\gamma_1^{(i)}+\gamma_0^{(i)}) \quad \mod \ 1 , 
\label{eq:Wr_broken}
\end{equation}
where $\Tw_i$ is the twist of the ribbon $R_A^{(i)}$, 
the angles $\gamma_1^{(i)}$ and $\gamma_0^{(i)}$ are determined by their trigonometric functions
\begin{eqnarray}
\cos\gamma_1^{(i)}={{\bf n}_{A1}^{(i)} \cdot {\bf t}_{A0}^{(j)} \over 
\sqrt{1-\left({\bf t}_{A0}^{(j)} \cdot {\bf t}_{A1}^{(i)}\right)^2}} , \quad
\sin\gamma_1^{(i)}={{\bf b}_{A1}^{(i)} \cdot {\bf t}_{A0}^{(j)} \over 
\sqrt{1-\left({\bf t}_{A0}^{(j)} \cdot {\bf t}_{A1}^{(i)}\right)^2}} , \label{eq:CSG1i} \\
\cos\gamma_0^{(i)}={-{\bf n}_{A0}^{(j)} \cdot {\bf t}_{A1}^{(i)} \over 
\sqrt{1-\left({\bf t}_{A0}^{(j)} \cdot {\bf t}_{A1}^{(i)}\right)^2}} , \quad
\sin\gamma_0^{(i)}={{\bf b}_{A0}^{(j)} \cdot {\bf t}_{A1}^{(i)} \over 
\sqrt{1-\left({\bf t}_{A0}^{(j)} \cdot {\bf t}_{A1}^{(i)}\right)^2}} , \label{eq:CSG2i}
\end{eqnarray}
and ${\bf t}_{A\upsilon}^{(i)}, {\bf n}_{A\upsilon}^{(i)}, {\bf b}_{A\upsilon}^{(i)}$
are the Frenet frames at the beginning ($\upsilon=0$) and at the end ($\upsilon=1$) 
of the $i$-th segment ($i=1,2,\ldots, n,\ \ j=1 + (i \mod n)$).

Note that the value of writhe generally depends on both the order of fragments and the orientations along them.

\subsection{A non-smoothly closed loop}
\label{nsmoothloop}

A particular case takes place when the segment $A$ forms a non-smoothly closed shape.
This means discontinuity of the tangent vector at the initial point and the tangent indicatrix is not closed. The whole procedure described above may be well applied to such a loop though one complication appears: the resulting closed curve to which the basic C\u{a}lug\u{a}reanu-White-Fuller formula is to be applied has a self-intersection point at the beginning of the loop considered. Generally, the writhe is not defined for such shapes. 
However, on the one hand, it was shown in Section \ref{existGauss} that the Gau\ss~integral exists 
unless the tangent at the loop starting point directs exactly opposite to the end tangent.
On the other hand, under the same limitation, we can restrict ourselves to consideration of
two limiting curves approaching the self-intersection shape from two different sides. 
As it is well known, the writhe jumps by $2$ as a curve crosses itself \cite{Fuller78}. 
Thus, the fractional part of the writhe is not affected by self-intersection and 
may be computed by Eq.~(\ref{eq:WRAS1}) or Eq.~(\ref{eq:WTgams}) in the same way as for the open segment.


\pagebreak
\section{Adding the writhe}

The aim of this section is to obtain an equation that expresses the writhe of 
a segment concatenated from two or more shorter segments as a function of the writhes of those smaller elements.


\subsection{A special case of a closed tangent indicatrix}

We start with a consideration of an open or closed segment 
$A = {\bf r}_A(s): [0, L_A] \rightarrow \R^3$ that has a (smoothly) closed
tangent indicatrix $\tilde{A}$ on ${\bf S}^2$. Let $D \in \tilde{A}$ be the point corresponding to the starting tangent ${\bf t}_A(0)$ and a point
$E \in \tilde{A}$ represents some other value of $s=s_1: 0 < s_1 < L_A$.
We draw  a geodesic $\tilde{G}$ to tie the points $E$ and $D$ (oriented from $E$ to $D$). Denote the two subsegments $A_1= {\bf r}_A(s): [0, s_1] \rightarrow \R^3$ and $A_2= {\bf r}_A(s): [s_1, L_A] \rightarrow \R^3$. Thus, $A_1+A_2=A$.
The writhe of the initial segment $A$ is, according to Eq.~(\ref{eq:WRAS}),
$$
\Wr_{A}=\frac{S_A}{2\pi} -1 \quad \mod \ 2 . 
$$
Apply now Eq.~(\ref{eq:WRAS1}) to the both parts of $A$:
$$
\Wr_{A_1}=\frac{S_{A_1+G}}{2\pi}  \quad \mod \ 1 , \quad
\Wr_{A_2}=\frac{S_{A_2-G}}{2\pi}  \quad \mod \ 1 . 
$$
(Recall that the sign ``$-$" in $-G$ denotes its reversed orientation.)

Clearly, $S_{A_1+G} + S_{A_2-G} = S_A$ and therefore
\begin{equation}
\Wr_{A}=\Wr_{A_1}+\Wr_{A_2} \quad \mod \ 1 .  \label{eq:wra12} 
\end{equation}

We can also reformulate this addition rule by expressing the writhe as a function of the twisting numbers. Then, Eqs.~(\ref{eq:WTgam}), (\ref{eq:WTgams}) imply
\begin{eqnarray}
\Wr_A=-\Tw_A  \quad \mod \ 1 , \nonumber \\
\Wr_{A_1}=-\Tw_{A_1} -\frac{\gamma_1 + \gamma_0}{2 \pi} \  \mod \ 1 , \ \
\Wr_{A_2}=-\Tw_{A_2} +\frac{\gamma_1 + \gamma_0}{2 \pi} \  \mod \ 1 . \nonumber
\end{eqnarray}
Here, $\gamma_1$ is an angle from ${\bf n}_A(s_1)$ to ${\bf n}_{G0}$, the initial normal vector of $G$. Respectively, $\gamma_0$ is an angle from ${\bf n}_{G1}$, the terminal normal vector of $G$, to ${\bf n}_A(0)$.

The twist $\Tw$ is additive, hence $\Tw_A=\Tw_{A_1}+\Tw_{A_2}$ and we again come to Eq.~(\ref{eq:wra12}).


\subsection{Another special case: two fragments}

\begin{figure}[htbp]
\begin{center}
\includegraphics[height=8cm]{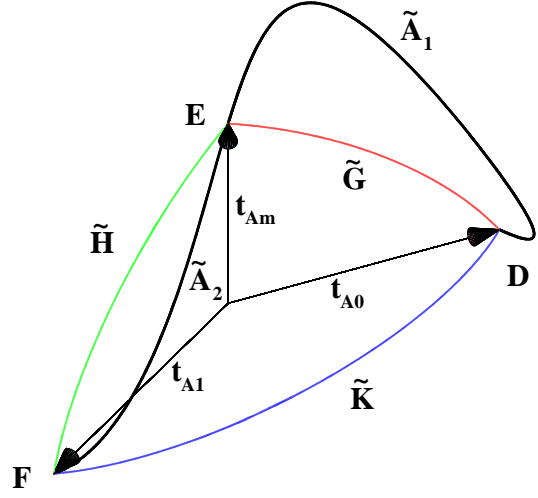}
\end{center}
\caption{The tangent indicatrix is divided by point $E$ into two fragments
$\tilde{A_1}$ and $\tilde{A_2}$; $\tilde{G}$, $\tilde{H}$, and $\tilde{K}$ are geodesics.}
\label{eq:add_writhe}
\end{figure}

If the segment $A$ forms a non-closed tangent indicatrix $\tilde{A} \in {\bf S}^2$, then the addition rule is more complex. Let $F \in \tilde{A}, F\neq D$ be the point corresponding to ${\bf t}_A(L_A)$. In addition to $\tilde{G}=ED$, we draw
two more geodesics: $\tilde{H}=FE$ and $\tilde{K}=FD$ (Fig.~\ref{eq:add_writhe}). Applying Eq.~(\ref{eq:WRAS1}) to the entire $\tilde{A}$$\tilde{A_1}$ and to its parts $\tilde{A_2}$ yields
\begin{eqnarray}
\Wr_{A}=\frac{S_{A+K}}{2\pi} \quad \mod \ 1 , \nonumber \\
\Wr_{A_1}=\frac{S_{A_1+G}}{2\pi}  \quad \mod \ 1 , \quad
\Wr_{A_2}=\frac{S_{A_2+H}}{2\pi}  \quad \mod \ 1 . \label{eq:GHK}
\end{eqnarray}

Taking into account the additivity of area, we have
\begin{equation}
S_{A+K}=S_{A_1+G}+S_{A_2+H}+S_{\triangle DEF} \quad \mod~4\pi . \label{eq:S}
\end{equation} 
The triangle $DEF$ is formed by geodesics and its signed area may be calculated as
\begin{equation}
S_{\triangle DEF} = \sigma_{DEF}(\angle DEF + \angle EFD + \angle FDE - \pi),
\label{eq:S2}
\end{equation} 
where 
$\sigma_{DEF}=\sign (({\bf t}_{A0} \times {\bf t}_{Am})\cdot{\bf t}_{A1})$, 
${\bf t}_{Am} \equiv {\bf t}_A(s_1)$.

The angles of the triangle can be found knowing the vectors of its vertices
$D,~E,~F$ 
(${\bf t}_{A0}, {\bf t}_{Am}$ and ${\bf t}_{A1}$, respectively): 
\begin{eqnarray}
\cos\angle DEF = {\cos\tilde{K} - \cos\tilde{G}\cos\tilde{H} \over
\sin\tilde{G} \sin\tilde{H}}  ,       \nonumber \\
\cos\angle EFD = {\cos\tilde{G} - \cos\tilde{H}\cos\tilde{K} \over
\sin\tilde{H} \sin\tilde{K}}  ,      \label{eq:cos} \\
\cos\angle FDE = {\cos\tilde{H} - \cos\tilde{K}\cos\tilde{G} \over
\sin\tilde{K} \sin\tilde{G}}  ,     \nonumber 
\end{eqnarray}
where
$$
\cos\tilde{G} = {\bf t}_{A0} \cdot {\bf t}_{Am} , \quad
\cos\tilde{H} = {\bf t}_{Am} \cdot {\bf t}_{A1} , \quad
\cos\tilde{K} = {\bf t}_{A1} \cdot {\bf t}_{A0} ,
$$
$$
\sin\tilde{G} = |{\bf t}_{A0} \times {\bf t}_{Am}| , \quad
\sin\tilde{H} = |{\bf t}_{Am} \times {\bf t}_{A1}| , \quad
\sin\tilde{K} = |{\bf t}_{A1} \times {\bf t}_{A0}| .
$$ 

Coming back to the addition rule for the writhe, we 
make use of Eq.~(\ref{eq:S}). The resulting equation then takes the form
\begin{equation}
\Wr_A = \Wr_{A_1} + \Wr_{A_2} + \frac{S_{\triangle DEF}}{2\pi} 
 \quad \mod~1 . \label{eq:W3}
\end{equation}  
By means of Eqs.~(\ref{eq:S2}), (\ref{eq:cos}), 
the area term in the last equation may be computed explicitly.

Note that if $E \in \tilde{K}$, then the area term in Eq.~(\ref{eq:W3}) 
is zero.


\subsection{The writhe of a growing curve}

Though Eq.~(\ref{eq:W3}) is valid modulo 1, it can be used for computation of the exact value of the writhe
of a curve which can be considered as being incrementally elongated.
Suppose we add to the curve $A_1$ a fragment $A_2$ which is short enough such that
$| \Wr_{A_2} + \frac{S_{\triangle DEF}}{2\pi}| \ll 1$.
Assume also that the exact value of $\Wr_{A_1}$ is known.
In the absence of self-intersections, the writhe is a continuous function of the arclength and we can
apply Eq.~(\ref{eq:W3}) omitting $\mod 1$ to get the exact value of the writhe
of the extended curve. Starting with a short fragment and reiterating the above procedure,
it is possible to compute the writhe of an arbitrarily long curve in $n$ steps where $n$
is the number of increments.


\subsection{The general case}

We now come to the general case. Let $A={\bf r}_A(s), 0 \le s \le L_A$
be a non-closed segment with the tangent indicatrix $\tilde{A} \in {\bf S}^2$.
Let $\{s_i\}, i=1,\ldots , N$, be the values of the arclength such that 
$0 < s_1 < s_2 < \ldots < s_i < s_{i+1} < \ldots < s_N < L_A$.
For every $s_i$ we can compute the tangent ${\bf t}_{Ai} \equiv {\bf t}_A(s_i)$
and denote ${\bf t}_{A0} = {\bf t}_A(0)$ and 
${\bf t}_{A,N+1} = {\bf t}_A(L_A)$. Let $E_i \in \tilde{A}$,
$i=0,\dots,N+1$, correspond to
${\bf t}_{Ai}$.

Choose some $j,~1 \le j \le N$, and apply the addition rule to the two segments
$E_0 E_j$ and $E_j E_{N+1}$:
$$
\Wr_{0,N+1}=\Wr_{0,j}+\Wr_{j,N+1}+\frac{S_{0,j,N+1}}{2\pi} \quad \mod~1 ,
$$ 
where $S_{0,j,N+1}$ is the signed area of the geodesic triangle $\triangle E_0 E_j E_{N+1}$.
If $j>1$, we can further apply the same rule to compute $\Wr_{0,j}$ as a sum of the writhes of subsegments. The same is also possible to do with $\Wr_{j,N+1}$,
if $j<N$. After iterating this procedure as far as possible (until all the $\Wr_{i,i+1},~i=0,\ldots,N$, are present in the summation formula) we finally
come to
\begin{equation}
\Wr_{0,N+1}=\sum_{i=0}^N \Wr_{i,i+1}+\frac{S_{0,1,\dots,N,N+1}}{2\pi} \quad \mod~1 ,
\label{eq:sumWr}
\end{equation}
where $S_{0,1,\dots,N,N+1}$ is the signed area of the spherical polygon formed by geodesics $E_{01},~E_{12},\dots,E_{j,j+1},\ldots,E_{N,N+1},~E_{N+1,0}$
(by $E_{ij}$ is denoted the geodesic $E_i E_j,~i\neq j$).
This area may be computed as a sum
$$
S_{0,1,\dots,N,N+1}=\sum_{i=1}^N S(\triangle E_0 E_i E_{i+1}) \quad \mod~4\pi .
$$
By using Eq.~(\ref{eq:S2}) the area of each spherical triangle can be presented in the form
$$
S(\triangle E_0 E_i E_{i+1}) = \sigma_{0,i,i+1}(\angle E_0 E_i E_{i+1} + 
\angle E_i E_{i+1} E_0 + \angle E_{i+1} E_0 E_i - \pi) ,
$$
$$
\sigma_{0,i,i+1}=\sign (({\bf t}_{A0} \times {\bf t}_{Ai}) \cdot {\bf t}_{Ai+1}) ,
$$
and each angle can be expressed as a function of ${\bf t}_{Ai}$ (see Eq.~(\ref{eq:cos})).

Another way to represent the area of a spherical $N+2$-gon is
$$
S_{0,1,\dots,N,N+1} = \sum_{i=0}^{N+1} \sigma_{i-1,i,i+1}(\angle E_{i-1} E_i E_{i+1} - \pi) \quad \mod~2\pi ,
$$
identifying $E_{-1}\equiv E_{N+1}$ and $E_{N+2}\equiv E_0$.


\section{The writhe of a polygonal line}

In this section we consider a curve consisting
of a number of straight line segments. Such a curve can serve as a model of a self-avoiding walk \cite{Paraped} or as a skeleton description of a linear polymer \cite{Chechetkin}. 

Though the polygonal line has a discontinuous tangent, it may be considered as a limit of a smooth curve which differs from the polygonal line by arbitrarily small planar arcs in the vicinities of every point of discontinuity. 
The consequitive tangent directions are then connected by geodesics on ${\bf S}^2$ \cite{Maggs}.
It can also describe the path of a scattered light beam \cite{MaggsScat}.

Conversely, an arbitrary smooth curve may be approximated by a polygonal line. In \cite{Cantarella}
one can find an estimate of the difference between the writhing numbers of a closed smooth curve
and a polygonal curve inscribed within.

Several methods have been proposed for computation of the writhe of closed polygons \cite{Klenin, Cimasoni}.
Here we give an exact expression for the fractional part of the writhe for an arbitrary polygonal line.
The formula does not involve the double summation.

Let $A=\{P_i\},~i=0,\ldots, N+1$, be a sequence of points in $\R^3$.
For every interval $(P_m,~ P_{m+1})$, a tangent vector is ${\bf t}_m=
\overrightarrow{P_m P_{m+1}} / |\overrightarrow{P_m P_{m+1}}|,~m=0,\ldots,N$.
We assume that $\overrightarrow{P_mP_{m+1}} \cap \overrightarrow{P_nP_{n+1}} = \emptyset,~ 
\mbox{if}\ {\bf t}_n \times {\bf t}_m = {\bf 0}, \ \forall m \neq n,~m,n=0,\ldots,N$.
The tangent indicatrix $\tilde{A} \in {\bf S}^2$ is a line consisting of fragments of geodesics $E_m E_{m+1},~m=0,\ldots,N-1,~E_m={\bf t}_m \cap {\bf S}^2 \in \tilde{A}$.
According to the above developed approach, we add the closing geodesic $E_N E_0$; it corresponds to an additional $(N+1)$-th fragment of $A$ such that ${\bf t}_{N+1} = {\bf t}_0$.

In order to make use of the formula for writhe, we have to calculate the signed area of a spherical polygon ${\cal P}=\{E_0\ldots E_N\}$. It may be represented as a union of triangles
$$
{\cal P}=\bigcup_{i=1}^{N-1} \triangle E_0 E_i E_{i+1} .
$$
Each triangle $\triangle E_0 E_i E_{i+1}$ has a signed spherical area
$$
S_{0,i,i+1}=  \sigma_{0,i,i+1} S(\triangle E_0 E_i E_{i+1}) \quad \mod~4\pi ,
$$
where $S(\triangle E_0 E_i E_{i+1}) = \alpha_{0,i,i+1}+\alpha_{i,i+1,0}+\alpha_{i+1,0,i}-\pi$
is a non-negative area of a spherical triangle with the angles $\alpha_{klm}$ and $\sigma_{klm}=\sign (({\bf t}_k \times {\bf t}_l) \cdot {\bf t}_m)$. The signed area of the polygon ${\cal P}$ is
$$
S_{\cal P} = \sum_{i=1}^{N-1} S_{0,i,i+1} \quad \mod~4\pi .
$$
The angles $\alpha_{klm}$ are functions of tangents such that
$$
\cos\alpha_{klm}=\frac{{\bf t}_k {\bf t}_m - ({\bf t}_k {\bf t}_l)({\bf t}_l{\bf t}_m)}
{|{\bf t}_k \times {\bf t}_l | |{\bf t}_l \times {\bf t}_m |} , \quad 0 \le \alpha_{klm} \le \pi .
$$

We now are able to rewrite Eq.~(\ref{eq:WRAS1}) in this particular case:
\begin{equation}
\Wr_A=\frac{1}{2\pi} \sum_{i=1}^{N-1} S_{0,i,i+1} \quad \mod \ 1 . \label{eq:F}
\end{equation}

This expression is valid for any curve in $\R^3$ that generates a tangent indicatrix containing only geodesic fragments. 
Also, it seems to be useful
 as an approximation while calculating the writhe of a smooth curve which can be properly discretized. 
The last means that the discretization should be better performed on its tangent indicatrix 
rather than on the curve in $\R^3$ itself. 
Conventional methods of integration may be applied to compute an approximate value of the area swept out on ${\bf S}^2$.


\section{The writhe and the Gau\ss~integral}

The writhe of the smooth closed curve may be expressed as the double integral (Eq. (\ref{eq:defWr})). It is evident that the writhe for an open segment of the length $L$ as defined above can no more be computed as the Gau\ss~integral over this segment though, in most cases, the double integral itself is also well defined for smooth non-closed curves.

Our aim here is to obtain a formula connecting the both values: on  the one hand, the writhe that relates to the difference between the linking number and the twisting for  the ribbon built with the geodesic closure and, on the other hand, simply the double integral taken over the open segment.


\subsection{An open curve and its closure}

Consider an open smooth non-self-intersecting curve $A = {\bf r}(s): [0, L] \rightarrow \R^3$. We assume here that the tangent vectors ${\bf t}(s)={\bf r}'(s)$ are neither parallel nor antiparallel at the ends: ${\bf t}(0)\neq\pm{\bf t}(L)$ (we will examine these cases later). We extend the curve $A$ with two straight line segments: $B : {\bf r}_B(s_1)={\bf r}(L)+s_1{\bf t}(L)$, $s_1\in [0, l]$ and $C : {\bf r}_C(s_2)={\bf r}(0)+s_2{\bf t}(0)$, $s_2\in [-l, 0]$.
Note that the both segments have the same length $l$. Now connect the end points of $B$ and $C$ with the straight line segment $D_l: {\bf r}_D (\xi)=(1-\xi){\bf r}_B(l)+\xi{\bf r}_C(-l)=
(1-\xi)({\bf r}(L)+l{\bf t}(L))+\xi({\bf r}(0)-l{\bf t}(0))$.
The direction of $D_l$ is determined by its tangent
$$
{\bf t}_D(l)=\frac{\frac{d{\bf r}_D(\xi)}{d\xi}}
{\left|\frac{d{\bf r}_D(\xi)}{d\xi}\right|}
=\frac{{\bf r}_C(-l)-{\bf r}_B(l)}{|{\bf r}_C(-l)-{\bf r}_B(l)|} .
$$

Now let the lengths of $B$ and $C$ increase to infinity and compute the limiting orientation of the tangent ${\bf t}_D$:
\begin{equation}
{\bf t}_{D\infty}=\lim_{l \to\infty} {\bf t}_D(l)=
-\frac{{\bf t}(L)+{\bf t}(0)}{|{\bf t}(L)+{\bf t}(0)|} . \label{eq:tD_inf}
\end{equation}

Thus, we see that, in the limit $l\to\infty$, $D_\infty$ lies in the plane
defined by the initial and end tangents of the segment $A$.

In the case when ${\bf t}(0)=\pm{\bf t}(L)$, we can also attach two straight line segments. If ${\bf t}(0)=-{\bf t}(L)$, then all straight lines connecting these segments belong to the same plane defined by ${\bf t}(0)$ and ${\bf r}(L)-{\bf r}(0)$. The case ${\bf t}(0)={\bf t}(L)$ requires a special consideration.

What we have now is a closed circuit $A+B+D_l+C$. It is smooth except for two points at the beginning and at the end of $D_l$. We modify $B$ and $D_l$ in the small vicinity of where they join themselves together by introducing a planar curvilinear segment $E$ with the tangent varying from ${\bf t}(L)$ to ${\bf t}_D$. All three segments involved belong to the same plane spanned by ${\bf t}(L)$ and ${\bf t}_D$. We can assume that the length of $E$ does not depend on $l$.
The length of the shortened segment $B_\ast$ is decreased to be $l_\ast$. 

The similar procedure may be carried out to smoothen the join of the segments $D_l$ and $C$. The new planar curvilinear segment $F$ belongs to the plane spanned by ${\bf t}(0)$ and ${\bf t}_D$.
The length of $F$ is the same for every $l$; without loss of generality, we assume that the length of the shortened segment $C_\ast$ equals $l_\ast$, as well.

We have come to the smooth closed curve $A+B_\ast+E+D_\ast+F+C_\ast$. We are interested in the limiting case when $l_\ast \to \infty$. The tangent indicatrix of the initial curve $A$ is closed then by a geodesic corresponding to the limiting curve $E_\infty+D_\infty+F_\infty$. 
This follows from the construction of these curves and from Eq.~(\ref{eq:tD_inf}).
The limiting curve $B_\infty+E_\infty+D_\infty+F_\infty$ may be considered as an implementation of the first part $G$ of the closure constructed in Section~\ref{arbitrary}. Thus, the writhe of the open segment $A$ may be computed as the writhe of the limiting closed curve $A+B_\infty+E_\infty+D_\infty+F_\infty+C_\infty$ and
its fractional part satisfies Eq.~(\ref{eq:WRAS1}) and Eq.~(\ref{eq:WTgams}).

However, for the smooth closed curve $A+B_\infty+E_\infty+D_\infty+F_\infty+C_\infty$, the writhe may be obtained by the double integral formula independently. Since the circuit consists of 6 parts, we are to consider all the pairs of them as they are involved in the double integration.
For the brevity, we will denote an integral over a pair of curves $P$ and $Q$ by $(P,Q)$. Clearly, $(P,Q)$ is the same as $(Q,P)$.

Before proceeding with this, we obtain some simple estimate of the value of the double integral
$$
I_2= \int\limits_0^{{\cal L}_2}\int\limits_0^{{\cal L}_1}I_{\Wr}(\sigma_1,\sigma_2) ~d\sigma_1 d\sigma_2 , 
$$
where
$$
I_{\Wr}(\sigma_1,\sigma_2)=\frac{({\bf r}_1(\sigma_1)-{\bf r}_2(\sigma_2))\cdot ({\bf t}_1(\sigma_1)\times{\bf t}_2(\sigma_2))}{|{\bf r}_1(\sigma_1)-{\bf r}_2(\sigma_2)|^3} .
$$

The integral $I_2$ is taken over two smooth curves ${\bf r}_1(\sigma_1), \sigma_1\in[0,{\cal L}_1]$
and ${\bf r}_2(\sigma_2), \sigma_2\in[0,{\cal L}_2]$. Let $\Delta \equiv \min\limits_{\sigma_1,\sigma_2} |{\bf r}_1(\sigma_1)-{\bf r}_2(\sigma_2)|>0$. Then
\begin{equation}
|I_2|\le \int\limits_0^{{\cal L}_2}\int\limits_0^{{\cal L}_1}|I_{\Wr}(\sigma_1,\sigma_2)| ~d\sigma_1 d\sigma_2 \le 
\int\limits_0^{{\cal L}_2}\int\limits_0^{{\cal L}_1}
\frac{d\sigma_1 d\sigma_2}{
|{\bf r}_1(\sigma_1)-{\bf r}_2(\sigma_2)|^2} \le \frac{{\cal L}_1 {\cal L}_2}{\Delta^2} .
\label{eq:estim}
\end{equation}

Eq.~(\ref{eq:estim}) implies that $\lim\limits_{\Delta \to \infty} I_2=0$ for any two curves of finite length. If one of the curves has its length of order $\Delta$ or less, i.e., ${\cal L}_i = {\cal O}(\Delta), i=1, 2$, and the other is of the finite length ${\cal L}_{3-i}$, then the integral $I_2$ vanishes as $\Delta \to \infty$, too.

We now come back to the integral over the pairs of curves. The integrals over $(B_\ast,B_\ast)$, $(B_\ast,E)$, $(B_\ast,D_\ast)$, $(E,E)$, $(E,D_\ast)$, $(D_\ast,D_\ast)$, $(D_\ast,F)$, $(D_\ast,C_\ast)$, $(F,F)$, $(F,C_\ast)$, $(C_\ast,C_\ast)$ equal zero because the integrand vanishes for coplanar curves. If ${\bf t}(0)=-{\bf t}(L)$, then the whole closure is planar and the integrals $(B_\ast,F)$, $(E,F)$, $(E,C_\ast)$ also vanish for every $l_\ast$. If ${\bf t}(0)\neq\pm{\bf t}(L)$, then the length of $D_{\ast}$ is of order $l_\ast$ for large $l_\ast$. Therefore, by applying Eq.~(\ref{eq:estim}), we obtain that the integrals $(B_\ast,F)$, $(E,F)$, $(E,C_\ast)$ as well as $(A,E)$, $(A,D_\ast)$, $(A,F)$ all approach zero as $l_\ast \to \infty$. We denote the remaining integrals that can be non-zero as follows:
$$
\Wy=\frac{1}{4\pi}\int\limits_A \int\limits_A I_{\Wr}(s,\tilde{s})~ds d\tilde{s},
$$
$$
\Sw_1=\frac{1}{2\pi}\int\limits_A \int\limits_{B_\infty} I_{\Wr}(s,s_1)~ds_1 ds, \quad
\Sw_2=\frac{1}{2\pi}\int\limits_A \int\limits_{C_\infty} I_{\Wr}(s,s_2)~ds_2 ds,
$$
$$
\Sq=\frac{1}{2\pi}\int\limits_{C_\infty} \int\limits_{B_\infty} I_{\Wr}(s_1,s_2)~ds_1 ds_2 .
$$
We call them ``the wrying'', ``the swirl'' and ``the squint'', respectively.

Thus, the writhe of $A$ may be represented as a sum
\begin{equation}
\Wr=\Wy+\Sw_1+\Sw_2+\Sq .    \label{eq:Main_eq}
\end{equation}

The first summand $\Wy$ is simply the double integral over the open segment under consideration. Therefore, Eq.~(\ref{eq:Main_eq}) provides a connection between this integral and the writhe.

Let us now examine the case of parallel tangents ${\bf t}(0)={\bf t}(L)$.
The two attached segments $B$ and $C$ have opposite directions then and, instead of the straight line $D$, we connect them by a circular arc $\breve{D}$ joining $B$ and $C$
at the same distance $l$ from the ends of $A$. The arc $\breve{D}$ lies in the plane spanned by the vectors
${\bf r}(L)-{\bf r}(0)$ and ${\bf t}(0)$ and its length is of order $l$ for large $l$. Again, the smoothening curves $E$ and $F$ can be constructed in the similar way as it is done in the regular case. Thus, we obtain the smooth planar closure of the curve $A$.

After letting the lengths of $B_\ast$ and $C_\ast$ go to infinity and analysing the double integral components in the expression for the writhe we come to the same Eq.~(\ref{eq:Main_eq}) with the right-hand terms defined as above.

It may occur that the ray $B$ or $C$ intersects the curve $A$. Then, generally, the writhe of the whole closed curve
$A+B+E+D+F+C$ is not determined. The situation is the same as for a non-smoothly closed loop 
(Section \ref{nsmoothloop}). In the generic case, when the tangents in the point of the intersection
are neither coincident nor of opposite direction, the fractional part of the writhe still can be found by the
examination of the two limiting positions of the curves in the vicinity of the intersection point. 
Since the writhe jumps by 2 as the curve goes through itself,
the half-sum of the writhes for those curves may be taken as the value of the writhe.
The same approach may be applied to another singular case when the rays $B$ and $C$ cross each other.
Moreover, the constraint of non-self-intersection of the initial open fragment $A$ my also be
weakened in the similar fashion.

Next we are going to clarify the structure of the integrals $\Sw$ and $\Sq$.


\subsection{The swirl}

Consider
\begin{eqnarray}
\Sw_1=\frac{1}{2\pi}\int\limits_{0}^L \int\limits_0^\infty \frac{({\bf r}(s)-{\bf r}(L)-s_1{\bf t}(L))\cdot ({\bf t}(s)\times{\bf t}(L))}{|{\bf r}(s)-{\bf r}(L)-s_1{\bf t}(L)|^3}~d s_1 ds = \nonumber \\
=\frac{1}{2\pi}\int\limits_0^L ({\bf R}(s)\cdot({\bf t}(s)\times{\bf t}_1) \hat{I}_{\Sw}(s)~ ds . \label{eq:Sw} 
\end{eqnarray} 
Here we denoted
$$
\hat{I}_{\Sw}(s) = \int\limits_0^\infty \frac{d s_1}{[({\bf R}-({\bf R}\cdot{\bf t}_1){\bf t}_1)^2
+({\bf R}\cdot{\bf t}_1-s_1)^2]^{\frac{3}{2}}} 
$$
and ${\bf R}\equiv{\bf R}(s)\equiv{\bf r}(s)-{\bf r}(L)$, ${\bf t}_1\equiv{\bf t}(L)$.

We can represent $\hat{I}_{\Sw}(s)$ as
$$
\hat{I}_{\Sw}(s)=\int\limits_0^\infty\frac{ds_1}{[a^2+(b-s_1)^2]^{\frac{3}{2}}} , \quad b\equiv b(s)\equiv{\bf R}\cdot{\bf t}_1, \quad a^2\equiv a^2(s)\equiv ({\bf R}-b {\bf t}_1)^2 ,
$$
and carry out the integration to get
\begin{equation}
\hat{I}_{\Sw}(s)=\left. 
\frac{s_1-b}{a^2\sqrt{a^2+(b-s_1)^2}}
\right|_0^\infty=
\frac{1}{\sqrt{a^2+b^2}(\sqrt{a^2+b^2}-b)} . \label{eq:ISw}
\end{equation}

Substituting $a(s), b(s)$ into Eq.~(\ref{eq:ISw}) and the result
 further into Eq.~(\ref{eq:Sw}) yields
\begin{equation}
\Sw_1=\frac{1}{2\pi}\int_0^L\frac{{\bf R}(s)\cdot({\bf t}(s)\times{\bf t}_1)}
{|{\bf R}(s)|(|{\bf R}(s)|-{\bf R}(s)\cdot{\bf t}_1)} ds .
\label{eq:Sw1}
\end{equation}

\begin{figure}[htbp]
\begin{center}
\includegraphics[height=8cm]{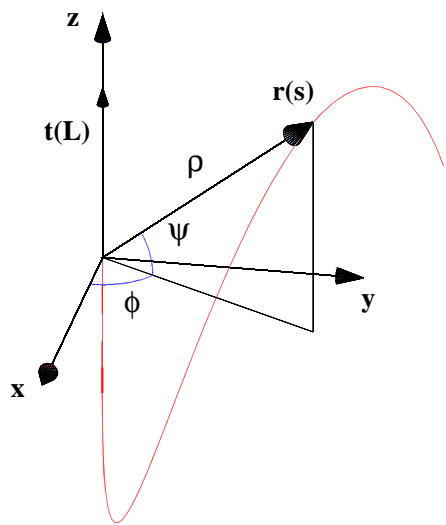}
\end{center}
\caption{The spherical coordinates $\rho, \phi, \psi$. The $z$-axis is directed along the tangent ${\bf t}(L)$ at the end point of the curve ${\bf r}(s)$.}
\label{eq:swirl}
\end{figure}

Let us introduce the spherical coordinate system with the origin at the point ${\bf r}(L)$ and let the $z$-axis be directed along the ray $B$ (Fig.~\ref{eq:swirl}). Then ${\bf R}(s)=(\rho \cos\psi \cos\phi, \rho \cos\psi \sin\phi, \rho\sin\psi)$, ${\bf t}_1=(0,0,1)$,
and $\rho\equiv\rho(s)$, $\phi\equiv\phi(s)$, $\psi\equiv\psi(s)$ are the functions describing the curve $A$.

In these coordinates, Eq.~(\ref{eq:Sw1}) takes the form
\begin{equation}
\Sw_1=\frac{1}{2\pi}\int\limits_0^L \phi' (1+\sin\psi) ds . \label{eq:Sw_sph}
\end{equation}
 
Note that the swirl is zero if $\phi(s)=const$, i.e., if
the whole curve $A$ is planar. The swirl also vanishes when $\psi(s)=-\frac{\pi}{2}$ which means that the curve $A$ is a straight line continuation of the ray $B$.

It is natural that the swirl is scale-invariant: it does not depend explicitly on 
how remote are the points of the curve from the ray.  

The second integral $\Sw_2$ over the ray $C_\infty$ has the same structure.


\subsection{The squint}

\begin{figure}
[htbp]    
\begin{center}
\includegraphics[height=8cm]{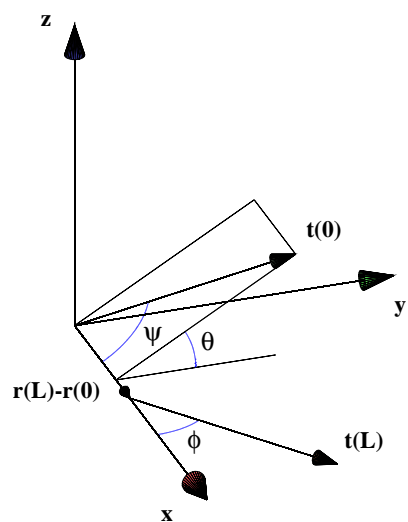}
\end{center}
\caption{The angles $\phi, \theta, \psi$ determine the orientation of the tangents ${\bf t}(0)$ and ${\bf t}(L)$ at the ends of the curve segment.}
\label{eq:trigon}
\end{figure}

Now consider the integral $\Sq$ over the two rays $B_\infty$ and $C_\infty$. It is convenient
to introduce the special Cartesian coordinates with the origin at the point ${\bf r}(0)$ and the $x$-axis directed along ${\bf r}(L)-{\bf r}(0)$ 
(Fig.~\ref{eq:trigon}). Let $y$-axis lie in the plane of the ray $B_\infty$ and the $z$-axis be chosen such that the whole coordinate system is right-handed. Denote by $\phi, \phi\in[0,\pi]$, the angle from the $x$-axis to the direction of ${\bf t}(L)$. 
The orientation of the ray $C_\infty$ is defined by two angles: $\psi, \psi\in[0,\pi]$, between the $x$-axis and ${\bf t}(0)$ and $\theta, \theta\in[0,2\pi]$,
between the $xy$-plane and the plane spanned by the $x$-axis and ${\bf t}(0)$.
In this coordinate system we may represent the both rays as follows:
\begin{eqnarray}
{\bf r}_B(s_1)=(g+s_1\cos\phi, s_1\sin\phi, 0), \quad g\equiv|{\bf r}(L)-{\bf r}(0)|,
\nonumber \\
\quad {\bf t}_B=(\cos\phi, \sin\phi, 0), \quad s_1\in[0,\infty], 
\nonumber \\
{\bf r}_C(s_2)=(s_2\cos\psi, s_2\sin\psi\cos\theta, s_2\sin\psi\sin\theta), \nonumber \\
{\bf t}_C=(\cos\psi,\sin\psi\cos\theta,\sin\psi\sin\theta), \quad s_2\in[-\infty,0].
\nonumber
\end{eqnarray}

We wish to compute the integral
\begin{eqnarray}
\Sq=\frac{1}{2\pi}\int\limits_{-\infty}^0 \int\limits_0^\infty \frac{({\bf r}_B(s_1)-{\bf r}_C(s_2))\cdot ({\bf t}_B(s_1)\times{\bf t}_C(s_2))}{|{\bf r}_B(s_1)-{\bf r}_C(s_2)|^3}~d s_1 ds_2 = \nonumber \\
=\frac{1}{2\pi}({\bf t}_B\times{\bf t}_C)\int\limits_{-\infty}^0 \int\limits_0^\infty \frac{{\bf r}_B(s_1)-{\bf r}_C(s_2)}{|{\bf r}_B(s_1)-{\bf r}_C(s_2)|^3}~d s_1 ds_2 = \nonumber \\
=\frac{g}{2\pi}\sin\psi\sin\phi\sin\theta\int\limits_{-\infty}^0 \int\limits_0^\infty
\frac{1}{(s_1^2+2 p s_1+q^2)^{\frac{3}{2}}}~d s_1 ds_2 , \nonumber
\end{eqnarray}
where $p\equiv p(s_2)\equiv g \cos\phi -s_2(\cos\psi\cos\phi+\sin\psi\sin\phi\cos\theta)$,
$q^2\equiv q^2(s_2)\equiv s_2^2 - 2 g s_2 \cos\psi +g^2$.

It is easy to perform the first integration:
\begin{eqnarray}
\Sq=\frac{g}{2\pi}\sin\psi\sin\phi\sin\theta\int\limits_{-\infty}^0 
\left(\left.\frac{s_1+p}{(q^2-p^2)\sqrt{s_1^2+2 p s_1+q^2}}\right|_0^\infty\right) d s_2 = \nonumber \\
=\frac{g}{2\pi}\sin\psi\sin\phi\sin\theta\int\limits_{-\infty}^0 
\frac{d s_2}{q(p+q)} = \nonumber \\
=\frac{g}{2\pi}\sin\psi\sin\phi\sin\theta\int\limits_{-\infty}^0 
\frac{1}{\sqrt{s_2^2 - 2 g s_2 \cos\psi +g^2}}\times \nonumber \\
\times\frac{d s_2}{
g \cos\phi -s_2(\cos\psi\cos\phi+\sin\psi\sin\phi\cos\theta)+\sqrt{s_2^2 - 2 g s_2 \cos\psi +g^2}} . \nonumber
\end{eqnarray}

The last integral can also be done and the result may be presented as an algebraic formula which does not depend on $g$, naturally.
However, the derivation of the final expression involves a complicated algebra and, instead, we prefer to obtain $\Sq$ in a different way.

We can consider the both rays and the straight line connecting them as an (infinite) polygonal line with three links. The writhe of this line, as defined in the previous section, is exactly equal to $\Sq$. Then the squint is essentially proportional to the signed area of the spherical triangle constituted by the geodesics that tie the vertices
corresponding to the vectors ${\bf t}(0)$, ${\bf r}(L)-{\bf r}(0)$, and ${\bf t}(L)$.

The triangle has its two sides equal to $\psi$ and $\phi$ and the angle between them $\theta$. By the side cosine theorem for spherical triangles, we find the third side $\chi$ from
$$
\cos\chi=\cos\psi\cos\phi+\sin\psi\sin\phi\cos\theta
$$
and the signed area of the triangle can be calculated by the generalized Heron's formula
$$
S=4 \nu \arctan\sqrt{\tan\frac{\Sigma}{2}\tan\frac{\Sigma-\phi}{2}\tan\frac{\Sigma-\psi}{2}\tan\frac{\Sigma-\chi}{2}},
$$
where $ \nu=\sign(({\bf r}(0)-{\bf r}(L))\cdot({\bf t}(0)\times{\bf t}(L)))=
\sign(\sin\psi\sin\phi\sin\theta)$ and
$\Sigma=\frac{1}{2}(\phi+\psi+\chi)$.

Then the squint is $\Sq=\frac{S}{2\pi}$.


\section{Example: helical shapes}

\label{sec_helix}

In this section we shall be dealing with a regular circular helix: ${\bf r}(s)=(\cos a s, \sin a s, \sqrt{1-a^2} s),~0 \le a \le 1,
~0 \le s \le L$, $s$ is the arc coordinate and $L$ the length of the segment. The limiting values of the parameter $a$ correspond to a straight line ($a=0$) and a circle of the unit radius ($a=1$). The curve is periodic with the period $T=\frac{2\pi}{a}$. The tangent, normal and binormal vectors are
$$
{\bf t}=(-a \sin a s, a \cos a s, \sqrt{1-a^2}) , \quad
{\bf n}=(-\cos a s, -\sin a s, 0) ,
$$
$$
{\bf b}=(\sqrt{1-a^2} \sin a s, -\sqrt{1-a^2} \cos a s, a) , 
$$   
and  $\kappa=a^2,~\tau=a\sqrt{1-a^2}$ are the curvature and the torsion,
respectively.

The tangent indicatrix of the helix is a circular arc on ${\bf S}^2$ of radius $a$.


\subsection{The writhe of an arbitrary segment of the helix}

We now illustrate the above developed approach by applying
Eq.~(\ref{eq:WTgams}) to the helix. We need to calculate the twist first. 
The normal ${\bf n}(s)$ is a well-defined continuous vector function and
\begin{equation}
\Tw(L)=\Tw_F(L)=\frac{1}{2\pi}\int\limits_0^L \tau ds = \frac{L}{2\pi}a\sqrt{1-a^2} . \label{eq:Twist}
\end{equation}

The second step is to calculate the angles $\gamma_1$ and $\gamma_0$. To this purpose, we find the tangent, normal and binormal vectors at the beginning ($s=0$) and at the end ($s=L$) points of the segment:
$$
{\bf t}_0=(0,a,\sqrt{1-a^2}),\quad {\bf n}_0=(-1,0,0),\quad {\bf b}_0=(0,-\sqrt{1-a^2},a),
$$
$$
{\bf t}_1=(-a\sin a L, a \cos a L, \sqrt{1-a^2}),\quad
{\bf n}_1=(-\cos a L, -\sin a L, 0),
$$
$$
{\bf b}_1=(\sqrt{1-a^2}\sin a L, -\sqrt{1-a^2}\cos a L, a) .
$$
It is possible now to make use of Eqs.~(\ref{eq:CSG1}),(\ref{eq:CSG2}) to obtain
\begin{eqnarray}
\cos{\gamma_1}=\cos{\gamma_0}=-\frac{\sigma\cos\frac{a L}{2}}
{\sqrt{1-a^2\sin^2\frac{a L}{2}}} , \label{eq:CG} \\
\sin{\gamma_1}=\sin{\gamma_0}=\frac{\sigma\sqrt{1-a^2}\sin\frac{a L}{2}}
{\sqrt{1-a^2\sin^2\frac{a L}{2}}} , \label{eq:SG}
\end{eqnarray}
where $\sigma=\sign\sin\frac{a L}{2}$.
Eqs.~(\ref{eq:CG}),(\ref{eq:SG}) imply $\gamma_1=\gamma_0$ and $\tan\gamma_1=\tan\gamma_0=-\sqrt{1-a^2}\tan\frac{a L}{2}$.

After substitution of these angles and the expression for $\Tw$
from Eq.~(\ref{eq:Twist}) into Eq.~(\ref{eq:WTgams}) we obtain the fractional part
of the writhe
$$
\Wr(L)=\frac{1}{\pi}\left[\arctan\left(\sqrt{1-a^2}\sin\frac{a L}{2},\cos\frac{a L}{2}\right)-
\frac{L}{2}a\sqrt{1-a^2}\right] \quad \mod~1 .
$$
We denote by $z=\arctan(x,y)$ a function such that
$\sin z = \frac{x}{\sqrt{x^2+y^2}}$, $\cos z = \frac{y}{\sqrt{x^2+y^2}}$, 
$-\pi< z \le \pi$.

Since the writhe of the helix should be continuous, we may rewrite the last equation in the form
\begin{eqnarray}
\Wr(L)=\frac{1}{\pi}\left[\arctan\left(\sqrt{1-a^2}\sin\frac{a L}{2},\cos\frac{a L}{2}\right)-
\frac{L}{2}a\sqrt{1-a^2}\right]+ \nonumber \\
+2\round\left(\frac{aL}{4\pi}\right) , \label{eq:WRH}
\end{eqnarray}
where $\round(x)$ is a function that gives an integer nearest to $x$.

\begin{figure}[htbp]    
\begin{center}
\includegraphics[height=8cm]{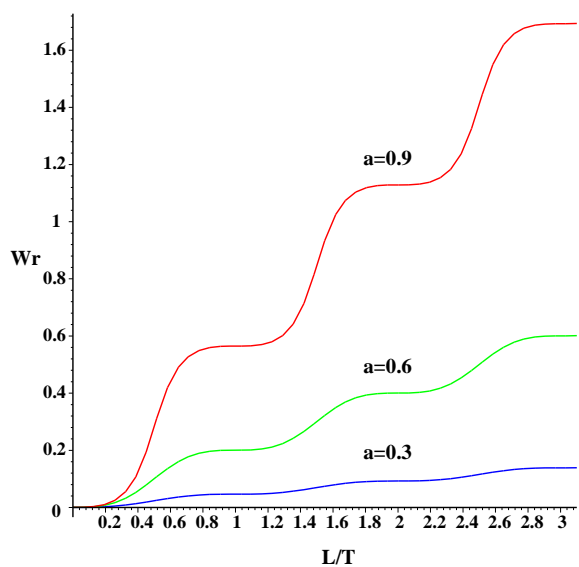}
\end{center}
\caption{The writhe $\Wr$ of the circular helix as a function of the arclength $L$ normalized on the period $T=2\pi/a$, for three different values of the parameter $a$.}
\label{eq:writhelix}
\end{figure}

The writhe $\Wr$ as a function of the arclength $L$ normalized on the period $T$ is presented in Fig.~\ref{eq:writhelix} for three different values of the parameter $a$.

We may use Eq.~(\ref{eq:WRH}) to compute the writhe for the particular lengths
$L=\frac{\pi m}{a}$ ($m$ is the number of half-periods):
\begin{equation}
\Wr\left(\frac{m T}{2}\right)=\frac{m}{2}(1-\sqrt{1-a^2}) \label{eq:mT}
\end{equation}
(cf. \cite{Fuller78, Ricca}).

In the limiting case $a \to 0$, Eq.~(\ref{eq:WRH}) produces $\Wr \to 0$
as it should be for a straight line. Another limiting value of the parameter is 
$1$: as $a \to 1$, the helix approaches a circle ${\bf r}(s)=(\cos s, \sin s, 0)$, covered $L/2\pi$ times. Then Eq.~(\ref{eq:WRH}) gives $\Wr \to \round\left(\frac{L}{2\pi}\right)$.


\subsection{The writhe by the double integral}

\label{sec_wrGint}

We wish to make use of the double integral formula (see Eq.~(\ref{eq:defWr}))
in order to compute the writhe of the infinite helix by an alternative means.
We take a helix $A$ with $2M$ turns, $M \ge 1$ being integer, 
and close it as described above with the planar curve $C$. 
(We could equally take an odd number of turns, but the even number facilitates the algebraic manipulations.)
We can build $C$ as a union of three fragments: $C=C_1+C_2+C_3$. All the fragments lie in the plane $x=1$. 
$C_1$ starts at the end of the helix in the point $(1,0,MT\sqrt{1-a^2})$ and it has the initial tangent $(0, a, \sqrt{1-a^2})$. Denote its length by $L_1$ and let it 
end at the point $(1, g, MT\sqrt{1-a^2})$, $g=const$, $g<L_1$, such that the whole curve is convex and
it lies in the half-plane $x=1,~z\ge MT\sqrt{1-a^2}$. The end tangent vector is $(0,0,-1)$.

The fragment $C_2$ is simply a straight line ${\bf r}_2(\sigma):(1,g,-\sigma)$, $\sigma\in
[-MT$ $\sqrt{1-a^2}, MT\sqrt{1-a^2}]$.

The last part $C_3$ is similar to $C_1$. It begins in the point $(1,g,-MT$ $\sqrt{1-a^2})$ with the tangent $(0,0,-1)$ and ends exactly in the beginning of the helix $(1,0,-MT\sqrt{1-a^2})$
having the tangent $(0,a,\sqrt{1-a^2})$. The convex curve $C_3$ has the length $L_3$ and it belongs to the half-plane $x=1, z\le-MT\sqrt{1-a^2}$.

It makes no sense to try to get the writhe for the infinitely lengthy helix. 
Instead, we consider the writhe per period defined as
\begin{equation}
\widetilde{\Wr}(T)= \lim_{M \to \infty} \frac{1}{2MT}\cdot
\frac{1}{4\pi}\int\limits_{A+C}\int\limits_{A+C}I_{\Wr}(s,\sigma) ds d\sigma , \label{eq:WRlim2}
\end{equation}
where
\begin{equation}
I_{\Wr}(s,\sigma)=\frac{({\bf r}(s)-{\bf r}(\sigma))\cdot ({\bf t}(s)\times{\bf t}(\sigma))}{|{\bf r}(s)-{\bf r}(\sigma)|^3} .
\label{eq:IWR}
\end{equation}

In Eq.~(\ref{eq:WRlim2}), we imply that the curves $C_1$ and $C_3$ remain unchanged as $M$ varies 
and only the length of the straight line fragment $C_2$ increases appropriately with $M$ growing. 
Then it may be shown that the right-hand side of Eq.~(\ref{eq:WRlim2}) depends on the shape of neither $C_1$ 
nor $C_3$.

The double integral in Eq.~(\ref{eq:WRlim2}) may be represented as a sum
$$
\int\limits_{A+C}\int\limits_{A+C}=\int\limits_{A}\int\limits_{A}+
2\sum\limits_{i=1}^3 \int\limits_{A}\int\limits_{C_i}
$$
(we took into account that 
$\int\limits_{A}\int\limits_{C_i}=\int\limits_{C_i}\int\limits_{A},~ i=1,2,3$,
and 
$\int\limits_{C_i}\int\limits_{C_j}=0,~\forall i,j=1,2,3$, since $I_{\Wr}\equiv 0$
at every point).

Now consider the mixed integral over $A$ and $C_1$. We split it into two
parts
\begin{equation}
\int\limits_{A}\int\limits_{C_1} I_{\Wr}\, d\sigma ds = \int\limits_{-MT}^{(M-1)T}\int\limits_0^{L_1}
I_{\Wr}\, d\sigma ds + \int\limits_{(M-1)T}^{MT}\int\limits_0^{L_1}
I_{\Wr}\, d\sigma ds .  \label{eq:int2}
\end{equation}
The second integral in Eq.~(\ref{eq:int2}) is taken over two finite fragments of a smooth curve and its value is finite, too,
and it does not depend on $M$. As to the first integral in Eq.~(\ref{eq:int2}),
we can estimate its numerator 
\begin{eqnarray}
|({\bf r}(s)-{\bf r}_1(\sigma))\cdot({\bf t}(s)\times{\bf t}_1(\sigma))|\le \nonumber \\
\le \sqrt{(L_1+2)^2+(L_1+\sqrt{1-a^2}(MT-s))^2}\le \nonumber \\
\le 2(L_1+1)+\sqrt{1-a^2}(MT-s) \nonumber
\end{eqnarray}
and the denominator
$$
|{\bf r}(s)-{\bf r}_1(\sigma))|^3\ge(1-a^2)^{\frac{3}{2}}(MT-s)^3 .
$$

Then
\begin{eqnarray}
\left| \int\limits_{-MT}^{(M-1)T}\int\limits_0^{L_1}I_{\Wr}\, d\sigma ds\right|\le
L_1\int\limits_{-MT}^{(M-1)T}\frac{2(L_1+1)+\sqrt{1-a^2}(MT-s)}
{(1-a^2)^{\frac{3}{2}}(MT-s)^3} ds = \nonumber \\
=\frac{L_1}{(1-a^2)^{\frac{3}{2}}}\int\limits_T^{2MT}\frac{2(L_1+1)+\sqrt{1-a^2} \varsigma}
{\varsigma^3} d\varsigma = \nonumber \\ =\frac{L_1}{(1-a^2)^{\frac{3}{2}}}\left[
(L_1+1)\left(\frac{1}{T^2}-\frac{1}{4M^2 T^2}\right)+\sqrt{1-a^2}\left(\frac{1}{T}-\frac{1}{2MT}\right)\right]\le \nonumber \\
\le\frac{L_1}{(1-a^2)^{\frac{3}{2}}}\left(\frac{L_1+1}{T^2}+\frac{\sqrt{1-a^2}}{T}\right) . \nonumber
\end{eqnarray}

Thus, we have shown that the mixed integral over $A$ and $C_1$ is bounded uniformly with respect to $M$. The same property for the integral over $A$ and $C_3$ can be proved similarly. We now deal with the third mixed integral over $A$ and $C_2$. It may be explicitly written as
$$
-a \int\limits_{-MT}^{MT}\int\limits_{-MT\sqrt{1-a^2}}^{MT\sqrt{1-a^2}}
\frac{1-\cos{as}-g\sin{as}}{(2+g^2-2\cos{as}-2g\sin{as}+(\sqrt{1-a^2} s +\sigma)^2)^{\frac{3}{2}}}~d\sigma ds .
$$ 

Since we are interested in obtaining the limit of the integral as $M \to \infty$,
we can consider
\begin{eqnarray}
I_2\equiv -a\int\limits_{-MT}^{MT}(1-\cos{as}-g\sin{as})\times \nonumber \\
\times\left(\int\limits_{-\infty}^\infty
\frac{d\sigma}{(2+g^2-2\cos{as}-2g\sin{as}+(\sqrt{1-a^2} s +\sigma)^2)^{\frac{3}{2}}}\right) ds .
\nonumber
\end{eqnarray}
It is easy to check that
$$
\int\limits_{-\infty}^\infty
\frac{d\xi}{(P+(Q+\xi)^2)^{\frac{3}{2}}}=\frac{2}{P},\quad P > 0 .
$$
Therefore,
$$
I_2=-2a\int\limits_{-MT}^{MT}\frac{1-\cos{as}-g\sin{as}}
{2+g^2-2\cos{as}-2g\sin{as}}~ ds .
$$ 
The last integral may be explicitly computed
\begin{eqnarray}
\int\frac{1-\cos{as}-g\sin{as}}{2+g^2-2\cos{as}-2g\sin{as}}~ ds = \nonumber \\
=\frac{s}{2}-\frac{1}{a}\arctan\left(\frac{1}{g^2}((4+g^2)\tan\frac{as}{2}-2g)\right)-\frac{\pi}{a}\round\left( \frac{as}{2\pi}\right) . \nonumber
\end{eqnarray}
It is periodic in $s$ with the period $T$ and, since
$$
\left|\frac{s}{2}-\frac{\pi}{a}\round\left(\frac{as}{2\pi}\right)\right|\le\frac{\pi}{2a}
$$
and
$$
\left|\frac{1}{a}\arctan\left(\frac{1}{g^2}((4+g^2)\tan\frac{as}{2}-2g)\right)\right|\le\frac{\pi}{2a} ,
$$
we get $|I_2|\le 4\pi$.

Summing up, we have proved that only double integration over the infinite helix $A$ may be retained in Eq.~(\ref{eq:WRlim2})
\begin{equation}
\widetilde{\Wr}(T)= \lim_{M \to \infty} \frac{1}{2MT}\cdot
\frac{1}{4\pi}\int\limits_{-MT}^{MT}\int\limits_{-MT}^{MT}I_{\Wr}(s,\sigma) ds d\sigma .
\label{eq:WRtilde}
\end{equation}

For the problem at hand, the integrand $I_{\Wr}$ may be reduced to 
\begin{eqnarray}
I_{\Wr}(s,\sigma)=a\sqrt{1-a^2}\times \nonumber \\
\times \frac{2\sin(\frac{a}{2}(s-\sigma))\left[
2\sin(\frac{a}{2}(s-\sigma))-a(s-\sigma)\cos(\frac{a}{2}(s-\sigma))\right]}
{\left[4\sin^2(\frac{a}{2}(s-\sigma))+(1-a^2)(s-\sigma)^2\right]^{\frac{3}{2}}} .
\nonumber
\end{eqnarray}
It depends only on the difference of the arc coordinates due to the translation invariance along the helix curve.
It is natural to make the change of variables $(s,\sigma)\rightarrow(u,v),~u=s-\sigma,~v=s+\sigma$ and to consider
$$
\tilde{I}_{\Wr}(u,v)=\tilde{I}_{\Wr}(u)=
a\sqrt{1-a^2}~\frac{2\sin(\frac{a}{2}u)\left[
2\sin(\frac{a}{2}u)-au\cos(\frac{a}{2}u)\right]}
{\left[4\sin^2(\frac{a}{2}u)+(1-a^2)u^2\right]^{\frac{3}{2}}} .
$$

For any fragment of the helix which length equals the period, the writhe is the same and it may be found as
\begin{equation}
\widetilde{\Wr}(T)= \lim_{M \to \infty} \frac{1}{2MT}\cdot
\frac{1}{4\pi}\int\limits_{-MT}^{MT} dv
\int\limits_{-MT}^{MT}
\tilde{I}_{\Wr}(u)du=\frac{1}{4\pi}\int\limits_{-\infty}^{\infty}\tilde{I}_{\Wr}(u)du .
\label{eq:WRlim1}
\end{equation}
The justification of why Eq.~(\ref{eq:WRtilde}) is equivalent to 
Eq.~(\ref{eq:WRlim1}) follows
from Lemma~2 (see Appendix~B) applied for $\varphi = 0$.

Consider an integral
$$
I_1=\int\limits_{-\infty}^{\infty}J_0(w)dw, \quad J_0(w)=\frac{\sin w (\sin w -w\cos w)}{(\sin^2 w + \lambda w^2)^{\frac{3}{2}}} ,
$$
where $\lambda=const, \lambda>0$ is a parameter; we put $\lambda=\frac{1-a^2}{a^2}$.

For small $w$,
\begin{equation}
J_0(w)=\frac{1}{3}\frac{|w|}{(1+\lambda)^{\frac{3}{2}}}+{\cal O}(w^3)=\frac{a^3}{3}|w|+
{\cal O}(w^3) . \label{eq:w3}
\end{equation}

The integral $I_1$ may be represented as a sum
$$
I_1=\int\limits_{-\infty}^{-\epsilon} J_0(w) dw +
\int\limits_{-\epsilon}^{\epsilon} J_0(w) dw +
\int\limits_{\epsilon}^{\infty} J_0(w) dw =
2\int\limits_{\epsilon}^{\infty} J_0(w) dw +
\int\limits_{-\epsilon}^{\epsilon} J_0(w) dw ,
$$
where $\epsilon > 0$ is small.

In view of Eq.~(\ref{eq:w3}), 
$$
\lim_{\epsilon \to 0}\int\limits_{-\epsilon}^{\epsilon} J_0(w) dw =0
$$
and we get
\begin{eqnarray}
I_1=2\lim_{\epsilon \to 0}\int\limits_{\epsilon}^{\infty} J_0(w) dw =
\left. 2\lim_{\epsilon \to 0}\frac{w}{\sqrt{\sin^2 w+\lambda w^2}}\right|_\epsilon^\infty= \nonumber \\
=2\left(\frac{1}{\sqrt{\lambda}}-\frac{1}{\sqrt{\lambda+1}}\right)=
2\frac{a(1-\sqrt{1-a^2})}{\sqrt{1-a^2}} . \nonumber
\end{eqnarray}

Returning to Eq.~(\ref{eq:WRlim1}) and making the identification $w=\frac{a}{2}u$,
we obtain
\begin{equation}
\widetilde{\Wr}(T)= \frac{a(1-\sqrt{1-a^2})}{2\pi}=\frac{\Wr(T)}{T} . \label{eq:Wrperiod}
\end{equation}
The last expression agrees with the value of the writhe for $m=2$ (Eq.~(\ref{eq:mT})).

We remark that the computation of the writhe (and twist) of the helix is
a favourite example of an application of the first Fuller theorem (e.g., see
\cite{Fuller78, Tyson, Ricca}) though the author has not come across the direct usage of the double integral formula in the literature. It is evident, that
the last approach is much more tedious, but it seems to be instructive 
to see it at work. To complete the picture, we show in the next section
how to calculate the same quantity by means of the Fuller second theorem.


\subsection{The writhe by the Fuller second theorem}

We consider two curves. The first one is a segment of a straight line $A_0:{\bf r}_0=(1,0,\sqrt{1-a^2} s)$ and the second is one period of the helix $A_1:{\bf r}_1=(\cos as, \sin as,\sqrt{1-a^2} s)$, the same as above. The common parameter $s$ varies from $0$ to $T=\frac{2\pi}{a}$ (note that it is not the arclength coordinate for $A_0$).

The Fuller second theorem is only applicable to closed curves, so we have to close both our segments. We shall do it in accordance with the procedure described above for periodic curves. Namely, we close $A_0$ and $A_1$ each with a curve
lying in the plane $x=1$ and consisting of three fragments. The closure of $A_0$
contains the parts:

\noindent
1) $C_{01}$, a semicircle of radius $R_0>0$ with the tangent ${\bf t}_{01}=(0,-\sin\phi$, $\cos\phi)$, 
$0\le \phi \le \pi, \phi$ is a new parameter;
 
\noindent
2) $C_{02}$, a segment of a straight line parallel to $A_0$:
${\bf r}_{02}=(1,2R_0$, $-\sqrt{1-a^2}$ $(T-\theta))$,
$\theta\in[0,T]$;

\noindent
3) $C_{03}$, a semicircle of radius $R_0$ with the tangent ${\bf t}_{03}=(0,-\sin\phi, \cos\phi)$, $\pi \le \phi \le 2\pi$.

The closing curve for the helix is built as follows:

\noindent
1) $C_{11}$, a semicircle of radius $R_1>\pi\sqrt{1-a^2}$ with the tangent
${\bf t}_{11}=(0$,$-\sin(\phi$ $+\phi_0), \cos(\phi+\phi_0)), ~0 \le \phi \le \pi,~
\phi_0$ is defined by $-\sin\phi_0=a,~ \cos\phi_0=\sqrt{1-a^2}$;

\noindent
2) $C_{12}$, a segment of a straight line:
${\bf r}_{12}=(1, a (a^2-1)\theta+2R_1\sqrt{1-a^2},$
$-(1-a^2)^{\frac{3}{2}} \theta +\frac{2\pi}{a}\sqrt{1-a^2} - 2R_1 a),~ \theta\in[0,T]$;

\noindent
3) $C_{13}$, a semicircle of radius $R_2=R_1-\pi\sqrt{1-a^2}$ with the tangent ${\bf t}_{13}=(0,-\sin(\phi+\phi_0), \cos(\phi+\phi_0)), \pi \le \phi \le 2\pi$.

At the joining points the tangent vector is not differentiable, but we may always make the closing curves smoother by small variations in the vicinities of the joining points. These modifications influence the result in no way.

The integrand in Eq.~(\ref{eq:Fuller2}) vanishes on the straight line segments of the closure and the rest two integrals over the semicircles cancel each other because ${\bf t}_{01}(\phi)=-{\bf t}_{03}(\phi+\pi)$ and 
${\bf t}_{11}(\phi)=-{\bf t}_{13}(\phi+\pi)$ for $0 \le \phi \le \pi$.

We now may forget about the closing segments and take the only integral over the initial range of parameter $s\in[0,T]$. The straightforward calculations yield the expression for the integrand function $\frac{a^3}{1+\sqrt{1-a^2}}=a(1-\sqrt{1-a^2})$ and we get again Eq.~(\ref{eq:Wrperiod}) for the writhe per period. In contrast to the previous section where we dealt with the infinite helix, we now have obtained the same formula by considering only one helical period.


\subsection{A double helix}

The procedure of the calculation of the writhe of a helix is very similar to that for a double helical shape.
The helix is assumed to be regular, circular and closed at both ends in the same manner as it is described above for open curves.
The difference is that each of the two closing curves joins two different helices.

In this section we show how to obtain the limiting value of the writhe per turn when the integer number of turns tends to infinity.
We include into our consideration the case when the double helix is not symmetric with respect to the central axis 
(as it takes place in the B-form of DNA, for example). Each of the two helical curves will be called a strand.

The first strand is described as the helix
$
D_0: {\bf r}_1(s_1)=(\cos a s_1$, $\sin a s_1$, $\sqrt{1-a^2} s_1),~0 \le a \le 1,
~S_0 \le s_1 \le S_1$, $s_1$ is the arc coordinate and $L=S_1-S_0$
the length of the segment (cf. the beginning of Section~\ref{sec_helix}). The second
strand may be obtained from $D_0$ by rotation through a (constant) offset angle $2\varphi$ around the central axis:
$ -D_2: {\bf r}_2(s_2)=(\cos (a s_2 + 2\varphi), \sin (a s_2 + 2\varphi), \sqrt{1-a^2} s_2),
~S_0 \le s_2 \le S_1$, $s_2$ is the arclength.
(Note that it is convenient to choose the same orientation for the parametrization
of both curves so that $-D_2$ is oriented against $D_0$.)

Let the double helix have $2M$ turns, $M$ integer (i.e. $L=2MT$, $T=\frac{2\pi}{a}$),
and $S_1=MT=-S_0$.
It is closed with two loops $D_1$ and $D_3$ of finite lengths $L_1$ and $L_3$,
respectively.
We assume that the loop $D_1$ is entirely contained in the cylinder $x_1^2+y_1^2 \le 2, z_1 \ge \sqrt{1-a^2} MT$ and
$D_3 \in \{x_3^2+y_3^2 \le 2, z_3 \le -\sqrt{1-a^2} MT\}$. 
The whole closed double helix may be represented as the sum
$D=D_0 + D_1 + D_2 + D_3$.

We wish to compute the writhe per period
\begin{equation}
\widetilde{\Wr}(T)= \lim_{M \to \infty} \frac{1}{2L+L_1+L_3}\cdot
\frac{1}{4\pi}\int\limits_{D}\int\limits_{D}I_{\Wr}(s,\sigma) ds d\sigma . \label{eq:WRlimD}
\end{equation}
where
 $I_{\Wr}(s,\sigma)$
 is as in Eq.~(\ref{eq:IWR}).

The double integral in Eq.~(\ref{eq:WRlimD}) may be represented as the sum
$$
\int\limits_{D}\int\limits_{D}=
\sum\limits_{i,j=0}^3 \int\limits_{D_i}\int\limits_{D_j}
 .
$$
Neither the integral $\int\limits_{D_1}\int\limits_{D_1}$ nor $\int\limits_{D_3}\int\limits_{D_3}$
depends on $M$, hence they do not affect the limit in Eq.~(\ref{eq:WRlimD}).
Consider the mixed integral over $D_0$ and $D_1$. We can split it similarly as it was done
for the single helix (cf. Eq.~(\ref{eq:int2}) 
\begin{equation}
\int\limits_{D_0}\int\limits_{D_1} I_{\Wr}\, d\sigma ds = \int\limits_{-MT}^{(M-1)T}\int\limits_0^{L_1}
I_{\Wr}\, d\sigma ds + \int\limits_{(M-1)T}^{MT}\int\limits_0^{L_1}
I_{\Wr}\, d\sigma ds .  \label{eq:intD2}
\end{equation}
The last double integral in Eq.~(\ref{eq:intD2}) exists
 and does not depend on $M$. 
As to the first integral in the right-hand side of Eq.~(\ref{eq:intD2}),
we can estimate its numerator 
\begin{eqnarray}
|({\bf r}_0(s)-{\bf r}_1(\sigma))\cdot({\bf t}_0(s)\times{\bf t}_1(\sigma))|\le \nonumber \\
\le\sqrt{(1+2)^2+(L_1/2+\sqrt{1-a^2}(MT-s))^2}\le 3+L_1/2+\sqrt{1-a^2}(MT-s) \nonumber
\end{eqnarray}
and the denominator
$$
|{\bf r}_0(s)-{\bf r}_1(\sigma))|^3\ge(1-a^2)^{\frac{3}{2}}(MT-s)^3 .
$$

Then
 it is easy to check that
$$
\left| \int\limits_{-MT}^{(M-1)T}\int\limits_0^{L_1}I_{\Wr}\, d\sigma ds\right|\le
\frac{L_1}{(1-a^2)^{\frac{3}{2}}}\left(\frac{L_1+6}{4T^2}+\frac{\sqrt{1-a^2}}{T}\right) . 
$$

We can conclude that the mixed integral over $D_0$ and $D_1$ is bounded uniformly with respect to $M$. 
The same is true
for the pairs $(D_3, D_0)$, $(D_1, D_2)$, $(D_2, D_3)$ and, of course, for $(D_1, D_3)$.
Now we see that the only pairs that count in Eq.~(\ref{eq:WRlimD})
are those involving both strands. Note that the integral $\int\limits_{D_0}\int\limits_{D_0} = \int\limits_{D_2}\int\limits_{D_2}$ was already
estimated as $\int\limits_{A}\int\limits_{A}$ (see Section \ref{sec_wrGint}).

The last integral to be computed is 
$\int\limits_{D_0}\int\limits_{D_2}=-\int\limits_{D_0}\int\limits_{-D_2}$. 
We apply Lemma~2 (see Appendix~B) to
obtain its limiting value which, according to Eq.~(\ref{eq:lem2}), will be
\begin{eqnarray}
-\frac{1}{4\pi}\int\limits_{-\infty}^{\infty}\tilde{I}_{\Wr,\varphi}(u) \ du = \nonumber \\
=-\left.\frac{\sqrt{1-a^2}}{4\pi}\cdot\frac{w}{\sqrt{\sin^2(w-\varphi)+\lambda w^2}}\right|_{-\infty}^{\infty} =
-\frac{\sqrt{1-a^2}}{2\pi}\cdot\frac{1}{\sqrt{\lambda}}=-\frac{a}{2\pi}, \nonumber \\
\varphi \neq 0, \quad w=\frac{au}{2}, \quad \lambda=\frac{1-a^2}{a^2}. \nonumber
\end{eqnarray}
Note that the result does not depend on $\varphi$.

Adding the above value to Eq.~(\ref{eq:Wrperiod}
) (which is actually the limiting value of the double integral
over the same helix), we finally obtain the writhe per period for the
double helix
\begin{equation}
\widetilde{\Wr}(T)= -\frac{a\sqrt{1-a^2}}{2\pi} .
\label{Wr2helix}
\end{equation}
It is necessary to clarify that the above formula gives the writhe normalized on the whole length
of both strands. Note also the negative sign of the writhe for the right-handed ($a>0$) double helix.

The symmetric case of $2\varphi=\pi$ was considered in \cite{Ricca}, where essentially the same formula for writhe was found.

We remark that the writhe for $n$-strand helical shapes ($n\ge3$) may be easily computed
on the basis of the results derived for the single and double helices.


\subsection{The writhe of the double helix of arbitrary length}

Here we continue dealing with the non-symmetric double helix, but now we consider the general case
when the helix does not need to have an integer number of turns. 
The length of each strand is the same and denoted by $L$.

We apply the equation for writhe that was found earlier for the sequence of disjoint fragments (Section~\ref{broken}).
In the double helix case, we have only two pieces: $A_1 \equiv D_0$ and $A_2 \equiv -D_2$, they are defined in the previous section, but now we set $S_0=0$ and $S_1=L$. The twist of each helix is the same
$$
\Tw_1=\Tw_2= \frac{L}{2\pi}a\sqrt{1-a^2} .
$$
(cf. Eq.~(\ref{eq:Twist}
)).

Now compute the vectors of the Frenet bases at all four ending points 
$A_0^{(1)}$, $A_1^{(1)}$, $A_0^{(2)}$, $A_1^{(2)}$:
\begin{eqnarray}
{\bf t}_{A0}^{(1)}=(0,a,\sqrt{1-a^2}),\quad {\bf n}_{A0}^{(1)}=(-1,0,0),
\quad {\bf b}_{A0}^{(1)}=(0,-\sqrt{1-a^2},a),
  \nonumber \\
{\bf t}_{A1}^{(1)}=(-a\sin a L, a \cos a L, \sqrt{1-a^2}),\quad
{\bf n}_{A1}^{(1)}=(-\cos a L, -\sin a L, 0),
  \nonumber \\
{\bf b}_{A1}^{(1)}=(\sqrt{1-a^2}\sin a L, -\sqrt{1-a^2}\cos a L, a), \nonumber \\
{\bf t}_{A0}^{(2)}=(a\sin(a L + 2\varphi), -a \cos(a L + 2\varphi), -\sqrt{1-a^2}),\nonumber \\
{\bf n}_{A0}^{(2)}=(-\cos(a L + 2\varphi), -\sin(a L + 2\varphi), 0),
  \nonumber \\
{\bf b}_{A0}^{(2)}=(-\sqrt{1-a^2}\sin(a L + 2\varphi), \sqrt{1-a^2}\cos(a L + 2\varphi), -a), \nonumber \\
{\bf t}_{A1}^{(2)}=(a\sin 2\varphi, -a \cos 2\varphi, -\sqrt{1-a^2}),\quad
{\bf n}_{A1}^{(2)}=(-\cos 2\varphi, -\sin 2\varphi, 0),
  \nonumber \\
{\bf b}_{A1}^{(2)}=(-\sqrt{1-a^2}\sin 2\varphi, \sqrt{1-a^2}\cos 2\varphi, -a) . \nonumber
\end{eqnarray}

It is easy to see that the angles $\gamma_0^{(i)}$, $\gamma_1^{(i)}$, $i=1,2$,
computed with the help of Eqs.~(\ref{eq:CSG1i}
), (\ref{eq:CSG2i}),
satisfy the following equations:
\begin{eqnarray}
\cos\gamma_1^{(1)}=-\cos\gamma_0^{(1)}=\cos\gamma_1^{(2)}=-\cos\gamma_0^{(2)}, \nonumber \\
\sin\gamma_1^{(1)}=\sin\gamma_0^{(1)}=\sin\gamma_1^{(2)}=\sin\gamma_0^{(2)}, \nonumber
\end{eqnarray}
which imply
$$
\gamma_1^{(1)} + \gamma_0^{(1)} + \gamma_1^{(2)} + \gamma_0^{(2)} = 0 \quad \mod \ 2\pi
$$
and according to Eq.~(\ref{eq:Wr_broken}) 
we have
$$
\Wr =-\frac{L}{\pi}a\sqrt{1-a^2} \quad \mod \ 1 .
$$
This expression is in compliance with the writhe per length for an integer number of turns (Eq.~(\ref{Wr2helix})).
If we think about the double helix as a continuously growing structure, 
then it is evident that the last equation gives not only the fractional part of $\Wr$, 
but its exact value as function of one strand length. 

Let $h$ be the length of the axis of the double helix, $h=L\sqrt{1-a^2}$.
Then
$$
\Wr = - \frac{h a}{\pi}
$$
We see that the growing double helix delivers an example of a family of curves $A(h)$, parametrized with the 
continuous parameter $h$, such that the writhe is a linear function of the length. 
Clearly, the writhe per (double helix) length is constant. Also, note that the writhe does not depend on the offset angle $\varphi$
which controls the mutual location of the strands (in terms of DNA we may reformulate the last
observation as an invariance property of writhe with respect to
the widths of the minor or major grooves).

\section{Conclusion}

We have been concerned with the generalization of the notion of the writhe
for an arbitrary space curve and with obtaining effective formulas for its computation.
We have analyzed various conditions on the position and orientation of the ends of the fragment in space starting with a periodic curve and finishing at the most  common case.

In all cases, an explicit construction of a closed ribbon was carried out. This makes it possible an application of the C\u{a}lug\u{a}reanu-White-Fuller equation to the determination of the writhe. A rule for the calculation of the writhe for a segment which is a union of smaller parts, is obtained. We examined a polygonal line as a special case.

A relation was established between the writhe and the Gau\ss~integral taken over the open fragment. The difference between
these two quantities may be represented as three single integrals.

The application of the formulas presented was demonstrated on the examplar curves, including regular single and double helices.
In particular, the writhe as a continuous function of arclength is defined for a regular helix.
It was shown that a double helix of finite length, with ``geodesic'' closures at the ends, 
provides an example of a one-parameter family of curves that realizes the linear dependence of the writhe on the length
and the writhe is invariant with respect to the value of the offset between the strands.
 
In Appendix~A, a new derivation of the formula connecting the writhe and the twist for a non-closed ribbon with one continuous edge is given.


\section{Appendix A: A formula for a non-closed ribbon}

The formula, we are about to deduce, relates to a ribbon based on a smooth closed curve but generated by a vector function which has a discontinuity. Such a ribbon may serve as a model to a nicked circular DNA, one strand of which is cleaved.

We will consider a closed smooth non-self-intersecting space curve
$A = {\bf r}(s): [0, L] \rightarrow \R^3$ of class $C^2$. We assume for the sake of simplicity that the parameter $s$ is the arclength. The smooth closure implies
${\bf r}(0)={\bf r}(L)$, ${\bf t}_0\equiv{\bf r}'(0)={\bf r}'(L)\equiv{\bf t}_1$,
${\bf r}''(0)={\bf r}''(L)$. 
Let the curve $A$ be equipped with the continuous vector function
${\bf u}(s): [0, L] \rightarrow {\bf S}^2$ such that ${\bf u}(s)\cdot{\bf r}'(s)=0, \forall s \in [0, L]$.

\begin{figure}[htbp]
\begin{center}
\includegraphics[height=8cm]{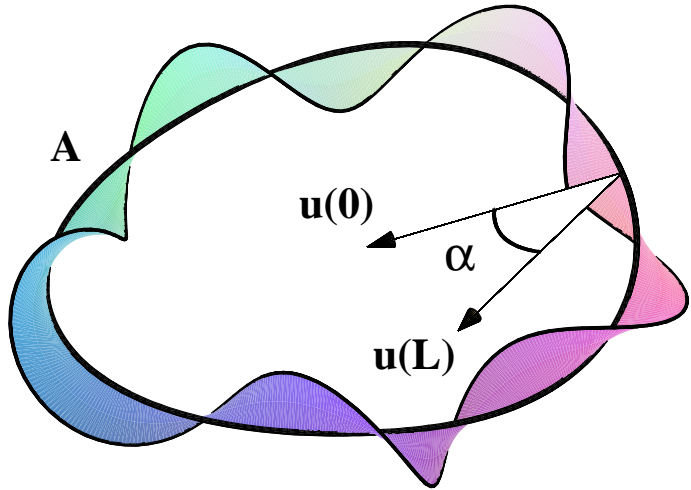}
\end{center}
\caption{A non-closed ribbon based on the smooth closed curve $A$.}
\label{eq:ribbon}
\end{figure}

In the general case, ${\bf u}(0)\neq {\bf u}(L)$ and we can define an angle $\alpha$ between these two vectors, measured up to modulo $2\pi$. This angle 
is a characteristic of discontinuity by which the ribbon $({\bf r}, {\bf u})$ fails to close (Fig.~\ref{eq:ribbon}).

Our aim here is to obtain an equation connecting the writhe of the curve $A$,
its twist and the angle $\alpha$.

Consider the tangent indicatrix $\tilde{A}$ of the curve $A$ on ${\bf S}^2$. It is a closed curve.
Choose a great circle plane containing ${\bf t}_0={\bf t}_1$. It has at least one common point with the tangent indicatrix. We look for a plane having at least two more intersections with $\tilde{A}$. Such a plane almost always exists. Indeed,
take a great circle which is tangent to $\tilde{A}$ at the initial point with
the tangent ${\bf t}_0$. Generically, there is an alternative: either the great circle intersects $\tilde{A}$ (the first case) or not (the second one).
In the first case, it is enough to rotate the great circle plane about ${\bf t}_0$ through a small angle to provide two additional points of intersections that
bifurcate from the initial one. Consider the second case. The whole tangent indicatrix of a \emph{closed} curve may not lie in only one of the two semispheres (see Theorem~1 in \cite{Vigodsky}). Then 
the closedness of $\tilde{A}$ implies at least two crossings in addition to the common initial point.
 
We should distinguish the singular case when the tangent indicatrix is itself 
a great circle. It corresponds to the planar space curve. We shall return to this case later.

For a regular configuration, we can choose two additional points $P_2: {\bf r}(s_2)$ and $P_3: {\bf r}(s_3)$ for which ${\bf t}_0+p {\bf t}_2 + q {\bf t}_3=0$, $p, q$ are real and ${\bf t}_2 \equiv {\bf r}'(s_2)$,
${\bf t}_3 \equiv {\bf r}'(s_3)$
(it is easy to show that the points may be always taken such that ${\bf t}_2 + {\bf t}_3 \neq 0$).

The starting point and $P_2$ and $P_3$ divide the curve $A$ into 3 parts, namely, $A=A_1+A_2+A_3$,
$A_1: {\bf r}(s), s\in [0,s_2]$, 
$A_2: {\bf r}(s), s\in [s_2,s_3]$,
$A_3: {\bf r}(s), s\in [s_3,L]$.

We now are ready to construct another curve by inserting 3 new segments of straight lines between the parts. The new curve $B$ consists of 6 fragments
$$
D_1: {\bf r}(0)+\sigma{\bf t}_0, \ \sigma \in [0,\epsilon_1];
$$
$$
B_1: {\bf r}(\sigma-\epsilon_1)+\epsilon_1{\bf t}_0, \ \sigma\in[\epsilon_1,\epsilon_1+s_2];
$$
$$
D_2: {\bf r}(s_2)+\epsilon_1{\bf t}_0+(\sigma-\epsilon_1-s_2){\bf t}_2, \
\sigma \in [\epsilon_1+s_2, \epsilon_1+s_2+\epsilon_2];
$$
$$
B_2: {\bf r}(\sigma-\epsilon_1-\epsilon_2)+\epsilon_1{\bf t}_0+\epsilon_2{\bf t}_2, \ \sigma\in[\epsilon_1+\epsilon_2+s_2,\epsilon_1+\epsilon_2+s_3];
$$
$$
D_3: {\bf r}(s_3)+\epsilon_1{\bf t}_0+\epsilon_2{\bf t}_2+(\sigma-\epsilon_1-\epsilon_2-s_3){\bf t}_3, \ \sigma\in[\epsilon_1+\epsilon_2+s_3,\epsilon_1+\epsilon_2+s_3+\epsilon_3];
$$
$$
B_3: {\bf r}(\sigma-\epsilon_1-\epsilon_2-\epsilon_3)+\epsilon_1{\bf t}_0+\epsilon_2{\bf t}_2+\epsilon_3{\bf t}_3, \ \sigma\in[\epsilon_1+\epsilon_2+\epsilon_3+s_3,\epsilon_1+\epsilon_2+\epsilon_3+L] .
$$
The compound curve $B(\epsilon_1, \epsilon_2, \epsilon_3)=D_1 + B_1+
D_2 + B_2+D_3 + B_3$ is closed if $\epsilon_1 {\bf t}_0+\epsilon_2{\bf t}_2+
\epsilon_3 {\bf t}_3=0$.  We put $\epsilon_2=p\epsilon_1$
and $\epsilon_3=q\epsilon_1$ and choose $\epsilon>0$ such that 
$B(\epsilon_1)$ would have no self-intersections for $\forall \epsilon_1, \epsilon_1\in[0,\epsilon]$. Thus, we have constructed an one-parameter family of closed pulled-out curves $B(\epsilon_1)$, $\epsilon_1\in[0,\epsilon]$, $B(0)=A$.

The next step is to define ribbons $(B(\epsilon_1),{\bf U}),~ \forall\epsilon_1\in(0,\epsilon]$. For the fragments $B_i, ~i=1,2,3$, we keep the same vectors ${\bf U}={\bf u}$ as for the respective parts $A_i,~i=1,2,3$. For the fragments $D_2$ and $D_3$, ${\bf U}={\bf u}(s_2)=const$
and ${\bf U}={\bf u}(s_3)=const$, respectively. For the remained fragment $D_1$, we
define ${\bf U}(\sigma),~\sigma\in[0,\epsilon_1]$ such that ${\bf U}(0)={\bf u}(L)$ and ${\bf U}(\epsilon_1)={\bf u}(0)$:
$$
{\bf U}(\sigma)={\bf u}(L)\cos\frac{\alpha}{\epsilon_1}\sigma+({\bf t}_0 \times {\bf u}(L))\sin\frac{\alpha}{\epsilon_1}\sigma .
$$
To find the angle $\alpha$ between ${\bf u}(0)$ and ${\bf u}(L)$ we
have the relationship
$$
{\bf u}(0)={\bf u}(L)\cos\alpha+({\bf t}_0 \times {\bf u}(L))\sin\alpha.
$$

The ribbon $(B(\epsilon_1),{\bf U})$ is continuous and the  
C\u{a}lug\u{a}reanu-White-Fuller formula can be applied to get
\begin{equation}
\Lk_B(\epsilon_1)=\Wr_B(\epsilon_1)+\Tw_B(\epsilon_1). \label{eq:CWF}
\end{equation}

Since we have chosen $\epsilon$ small enough to exclude self-intersections 
of the curve (more accurately, we have to require an absence
of self-crossings for the whole ribbon), the right-hand side of Eq.~(\ref{eq:CWF})
does not depend on $\epsilon_1$.

By the construction, the tangent indicatrix of $B(\epsilon_1)$ is the same for any $\epsilon_1$ (namely, it is $\tilde{A}$ because $B(0)=A$).
Therefore, due to the Fuller first theorem (and taking into account the absence of self-intersections), $\Wr_B(\epsilon_1)=\Wr_A$.

The twist of $B(\epsilon_1)$ may be represented as a sum
$$
\Tw_B(\epsilon_1)=\Tw_{D1}(\epsilon_1)+\Tw_{B1}+\Tw_{D2}(\epsilon_1)+\Tw_{B2}+\Tw_{D3}(\epsilon_1)+\Tw_{B3} .
$$
Clearly, $\Tw_{D2}(\epsilon_1)\equiv 0$ and $\Tw_{D3}(\epsilon_1)\equiv 0$,
because the vector ${\bf U}$ is constant on these straight-line fragments. Further, $\Tw_{B1}+\Tw_{B2}+\Tw_{B3}=\Tw_A$, the twist of the initial non-closed ribbon
(which is well defined).

The twist of the remaining part $D_1$ may be readily computed
$$
\Tw_{D1}=\frac{\alpha}{2\pi}
$$
and it does not depend on $\epsilon_1$.

We now can rewrite Eq.~(\ref{eq:CWF}) as
$$
\Lk_B=\Wr_A+\Tw_A+\frac{\alpha}{2\pi} .
$$
We may interpret the last equation as a definition of the linking number for the non-closed ribbon $A$ (cf. the last equation in \cite{Maddocks} and Eq.~(7.2) in \cite{Fuller78}):
$$
\Lk_A=\Wr_A+\Tw_A, \quad \Lk_A=\Lk_B-\frac{\alpha}{2\pi} .
$$

In conclusion, we explain how one has to treat the degenerated case of a planar curve $A$. Clearly, two insertions are sufficient then. Indeed, we can choose the second point on $\tilde{A}$ as diametrically opposed to the starting one, i.e., ${\bf t}_0+{\bf t}_2 = 0$. In other words, we can proceed the same way as in the regular case after putting $p=1$ and $q=0$. It is evident, that the result will not change.


\section{Appendix B: Estimation of integrals}

In this section two lemmas are proved that are useful for computation of the Gau\ss~integral
for regular helices.

\noindent {\bf Lemma~1.} {\it Let 
$$
J_\theta(w) = \frac{\sin(w-\theta)[\sin(w-\theta) - w \cos(w-\theta)]}{[\sin^2(w-\theta)+\lambda w^2]^{3/2}} , 
$$

where $\lambda$, $\theta$ are parameters and $\lambda > 0$.

Then the integral
$$
I_{\theta\infty}=\int\limits_0^\infty w J_\theta(w) dw
$$
exists.
} 

Consider an integral
$$
I_{\theta}(w_0)=\int\limits_0^{w_0} w J_\theta(w) dw
$$
and integrate it by parts using the equality
$$
\int\limits_0^{w_0} J_\theta(w) dw = \frac{w}{\sqrt{\sin^2(w-\theta)+\lambda w^2}} .
$$
Then
\begin{eqnarray}
I_{\theta}(w_0)=\frac{w_0^2}{\sqrt{\sin^2(w_0-\theta)+\lambda w_0^2}} - \hat{I}(w_0), \nonumber \\ 
\hat{I}(w_0)=\int\limits_0^{w_0} \frac{w}{\sqrt{\sin^2(w-\theta)+\lambda w^2}}  dw . \nonumber
\end{eqnarray}
Now we estimate $\hat{I}(w_0)$ for $w_0 > 0$. On the one hand,
$$
\hat{I}(w_0) < \int\limits_0^{w_0} \frac{d w}{\sqrt{\lambda}} =  \frac{w_0}{\sqrt{\lambda}}
$$
and on the other,
$$
\hat{I}(w_0) > \int\limits_0^{w_0} \frac{w d w}{\sqrt{1+\lambda w^2}} =  \frac{\sqrt{1+\lambda w_0^2} - 1}{\lambda} .
$$
Therefore, we have
\begin{eqnarray}
\frac{w_0^2}{\sqrt{\sin^2(w_0-\theta)+\lambda w_0^2}} - \frac{w_0}{\sqrt{\lambda}} < I_{\theta}(w_0) < \frac{w_0^2}{\sqrt{\sin^2(w_0-\theta)+\lambda w_0^2}} - \nonumber \\ 
-\frac{\sqrt{1+\lambda w_0^2} - 1}{\lambda} . \nonumber
\end{eqnarray}
Letting $w_0$ go to the infinity in the last inequality leads to
$$
0 < I_{\theta\infty} = \lim_{w_0 \to \infty} I_{\theta}(w_0) < \frac{1}{\lambda}
$$
which proves the statement of the lemma.

\noindent {\bf Lemma~2.} {\it Let 
\begin{eqnarray}
I_{\Wr,\varphi}(s,\sigma)=a\sqrt{1-a^2}\times \nonumber \\
\times \frac{2\sin(\frac{a}{2}(s-\sigma)-\varphi)\left[
2\sin(\frac{a}{2}(s-\sigma)-\varphi)-a(s-\sigma)\cos(\frac{a}{2}(s-\sigma)-\varphi)\right]}
{\left[4\sin^2(\frac{a}{2}(s-\sigma)-\varphi)+(1-a^2)(s-\sigma)^2\right]^{\frac{3}{2}}} .
\nonumber
\end{eqnarray}

Then 
\begin{equation}
\widetilde{\Wr}(T)= \lim_{M \to \infty} \frac{1}{2MT}\cdot
\frac{1}{4\pi}\int\limits_{-MT}^{MT}\int\limits_{-MT}^{MT}I_{\Wr,\varphi}(s,\sigma) ds d\sigma = 
\frac{1}{4\pi}\int\limits_{-\infty}^{\infty}\tilde{I}_{\Wr,\varphi}(u) du ,
\label{eq:lem2}
\end{equation}
where
$$
\tilde{I}_{\Wr,\varphi}(u)=
 a\sqrt{1-a^2}~\frac{2\sin(\frac{a}{2}u-\varphi)\left[
2\sin(\frac{a}{2}u-\varphi)-au\cos(\frac{a}{2}u-\varphi)\right]}
{\left[4\sin^2(\frac{a}{2}u-\varphi)+(1-a^2)u^2\right]^{\frac{3}{2}}} .
$$
} 

First of all, note that $I_{\Wr,\varphi}(s,\sigma)=I_{\Wr,-\varphi}(-s,-\sigma)$ 
(and $\tilde{I}_{\Wr,\varphi}(u)=\tilde{I}_{\Wr,-\varphi}(-u)$, as well).
Next, we make the change of variables $(s,\sigma)\rightarrow(u,v),~u=s-\sigma,~v=s+\sigma$ to
obtain
\begin{eqnarray}
\int\limits_{-MT}^{MT}\int\limits_{-MT}^{MT}I_{\Wr,\varphi}(s,\sigma) ds d\sigma = 
\frac12\left[\int\limits_{-MT}^{MT} \int\limits_{-MT}^{MT} \tilde{I}_{\Wr,\varphi}(u) du dv - \right. \label{eq:i02} \\
 - 2 \int\limits_{0}^{MT} \int\limits_{0}^{u} (\tilde{I}_{\Wr,\varphi}(u) + \tilde{I}_{\Wr,-\varphi}(u)) dv du + \nonumber \\
\left.+ 2 \int\limits_{MT}^{2MT} \int\limits_{0}^{2MT-u} (\tilde{I}_{\Wr,\varphi}(u) + \tilde{I}_{\Wr,-\varphi}(u)) dv du \right]. 
\nonumber 
\end{eqnarray}

Due to Lemma~1,
$$
\lim_{M \to \infty} \frac{1}{MT} \int\limits_{0}^{MT}\int\limits_{0}^{u} \tilde{I}_{\Wr,\pm\varphi}(u) dv du=
\lim_{M \to \infty} \frac{1}{MT} \int\limits_{0}^{MT} u \tilde{I}_{\Wr,\pm\varphi}(u) du = 0 .
$$

It is also evident that
\begin{eqnarray}
\lim_{M \to \infty} \frac{1}{MT} \int\limits_{MT}^{2MT} \int\limits_{0}^{2MT-u} \tilde{I}_{\Wr,\pm\varphi}(u) dv du =
\nonumber \\
=\lim_{M \to \infty} \frac{1}{MT} \int\limits_{MT}^{2MT} (2MT-u) \tilde{I}_{\Wr,\pm\varphi}(u) du = 0 ,\nonumber
\end{eqnarray}
because $\tilde{I}_{\Wr,\pm\varphi}(u) = {\mathcal O}(u^{-2})$ for $u \to\infty$.

Therefore, we may consider only the limit of the first pair of the integrals in the right-hand side of 
Eq.~(\ref{eq:i02}) which will be
$$
\lim_{M \to \infty} \frac{1}{2MT}\cdot
\frac{1}{4\pi}\int\limits_{-MT}^{MT}\int\limits_{-MT}^{MT} \tilde{I}_{\Wr,\varphi}(u) du dv = \frac{1}{4\pi}\int\limits_{-\infty}^{\infty}\tilde{I}_{\Wr,\varphi}(u) du .
$$

\section*{Acknowledgements}

Most of this work was carried out during the author's stay at the Institute of Technical Mechanics of the Technical University of Karlsruhe. 
The support from the Alexander von Humboldt Foundation is gratefully acknowledged.
The author would like to express his thanks to Jens Wittenburg for his hospitality and attention to the work.

The author is also pleased to thank John Maddocks for helpful discussions and his support.


\end{document}